\documentclass[final,5p,times]{elsarticle}

\usepackage{amsmath, amssymb, mathtools}
\usepackage{amsthm}

\usepackage{booktabs}       
\usepackage{array}          
\usepackage{tabularx}       
\usepackage{longtable}      
\usepackage{makecell}       
\newcommand{\cmark}{\checkmark}
\newcommand{\xmark}{--}

\usepackage{graphicx}
\usepackage{subcaption}
\usepackage{caption}

\usepackage{indentfirst}
\usepackage{balance}        

\usepackage{hyperref}
\hypersetup{
    colorlinks=true,
    linkcolor=blue,
    filecolor=magenta,
    urlcolor=cyan,
}

\usepackage{tikz}
\usetikzlibrary{positioning}

\usepackage{pifont}         
\usepackage{xspace}
\usepackage{listings}
\usepackage[table]{xcolor}
\rowcolors{3}{gray!6}{white}

\journal{---}

\begin{document}

\begin{frontmatter}

\title{Sliding-Mode Control Strategies for PMSM speed control: A Comprehensive Review, Taxonomy and Research Gaps}

\author[1]{Abdullah Ajasa}
\author[1,3]{Mubarak Badamasi Aremu}
\author[1,2,3,*]{Ali Nasir} \ead{alinasir@kfupm.edu.sa}

\affiliation[1]{organization={Control and Instrumentation Engineering Department, King Fahd University of Petroleum \& Minerals (KFUPM)}, 
            city={Dhahran}, 
            country={Saudi Arabia}}

\affiliation[2]{organization={Interdisciplinary Center for Aviation and Space Exploration (Guest Scholar), King Fahd University of Petroleum \& Minerals (KFUPM)},
            country={Saudi Arabia}}

\affiliation[3]{organization={Interdisciplinary Research Center (IRC) for Intelligent Manufacturing and Robotics, King Fahd University of Petroleum \& Minerals},
            country={Saudi Arabia}}

\affiliation[*]{Corresponding author: Ali Nasir (email: ali.nasir@kfupm.edu.sa)}

\begin{abstract}
Permanent Magnet Synchronous Motors (PMSMs) are widely employed in high-performance drive systems due to their high efficiency, power density, and precise dynamic behavior. However, nonlinearities, load disturbances, and parameter uncertainties present persistent challenges to control. Sliding-Mode Control (SMC) remains one of the most reliable strategies for high-performance PMSM drives. Yet, the rapid proliferation of adaptive, fractional-order, and intelligent variants has fragmented recent literature. This paper presents a comprehensive review and taxonomy of SMC-based PMSM speed-control methods published between 2020 and 2025.  More than 200 studies are systematically analyzed and classified according to control order, surface design, disturbance-observer integration, optimization approach, and intelligent augmentation.  Trends in publication activity,
dominant hybrid structures, and application domains are quantitatively summarized.  The review
reveals a clear evolution from conventional discontinuous SMC toward adaptive, higher-order,
and data-driven frameworks that mitigate chattering while preserving robustness.  Persistent
research gaps are identified in hardware validation, energy-efficiency assessment, and
real-time tuning strategies.  The taxonomy and critical synthesis provided herein establish
a coherent reference for researchers and form the conceptual foundation for the companion
paper (Part II), which delivers a unified benchmark and comparative simulation study of
representative SMC designs.
\end{abstract}


\end{frontmatter}

\section{Introduction} \label{sec: intro} 
Permanent Magnet Synchronous Motors (PMSMs) are among the most widely used electrical machines in advanced industrial systems due to their high efficiency, compact structure, and superior torque-to-inertia ratio \cite{podmiljvsak2024future}. They feature permanent magnets mounted either on the surface or embedded within the rotor (i.e., surface-mounted or interior permanent magnet synchronous motors, or PMSMs) \cite{lindh2009comparison}, producing a constant rotor flux without the need for excitation current. This design results in reduced losses and high power density \cite{pillay1989modeling}. PMSMs are extensively deployed in electric vehicles, robotics, precision manufacturing, renewable energy systems, and aerospace applications, where precise torque and speed control are critical \cite{karboua2023robust,zhao2023review,ozcciflikcci2024overview,azom2025recent,azom2025challenges}. Their combination of performance and compactness makes them ideal for applications with tight space and energy constraints. However, despite their growing adoption, PMSMs present significant control challenges that limit performance in real-world scenarios.

Strong nonlinearities, parameter uncertainties, and sensitivity to external disturbances characterize the dynamics of PMSMs. Factors such as load torque variations, magnetic saturation, temperature-dependent resistance drift, and back EMF coupling make the control task highly nontrivial \cite{nicola2020sensorless}. Moreover, high-performance PMSM control systems often require accurate position or speed feedback, which introduces sensor-related issues, including noise, cost, and reliability concerns \cite{gieras2018linear}. These challenges necessitate the use of robust, high-bandwidth control strategies that can ensure stability and performance across a wide range of operating conditions \cite{sakunthala2018review}.

Traditional control methods, such as Proportional-Integral-Derivative (PID) controllers and Field-Oriented Control (FOC), have been widely adopted in industrial Permanent Magnet Synchronous Motor (PMSM) drives due to their simplicity and ease of implementation \cite{zheng2011design}. In particular, Proportional-Integral (PI) controllers are commonly used for speed loop regulation because of their intuitive tuning and low computational overhead \cite{zhang2011direct}. However, their performance deteriorates significantly in the presence of nonlinearities, parameter uncertainties, and external disturbances. Notable limitations include excessive overshoot, poor robustness, strong dependency on accurate system modeling, and limited disturbance rejection capabilities. The increasing demand for high-performance PMSM drives, characterized by fast dynamic response, precise tracking, and resilience to system uncertainties, has driven the development of more advanced control strategies \cite{sakunthala2018review}. One persistent challenge in PMSM control is the requirement for rotor position information, which is typically obtained via position or speed sensors such as magnetic resolvers or optical encoders. While effective, these sensors introduce several issues, including increased cost, sensitivity to noise, reduced reliability in harsh environments, and added system complexity.

Moreover, torque regulation in PMSMs is typically achieved by controlling the stator (armature) current, as electromagnetic torque is directly proportional to it \cite{zhong1999direct,frikha2023multiphase}. This has motivated the exploration of nonlinear and robust control techniques that can cope with the motor’s nonlinear behavior and sensitivity to operating conditions. Several approaches have been proposed in recent literature, including adaptive backstepping control for PMSMs \cite{wang2019adaptive,li2019model}, model predictive control (MPC) strategies for current loop optimization \cite{zhang2017performance,djouadi2024improved}, and intelligent controllers based on neural networks and fuzzy logic \cite{sakunthala2017study,nouaoui2024speed,mohajerani2024neural}. Despite these developments, achieving optimal control performance under real-world conditions remains challenging for conventional linear and gain-scheduled controllers due to the PMSM’s strong nonlinearities, coupled dynamics, and external disturbances. This has led to growing interest in SMC, which offers inherent robustness, disturbance rejection, and real-time adaptability to system uncertainties \cite{weijie2014sliding}.

Sliding Mode Control (SMC) is a widely studied nonlinear control strategy known for its inherent robustness against system uncertainties and external disturbances \cite{gambhire2021review}. At its core, SMC operates by enforcing system trajectories onto a predefined sliding surface, where the closed-loop dynamics become insensitive to matched uncertainties. This is achieved through a discontinuous control law that drives the system to the sliding surface and maintains it there, ensuring fast convergence and strong disturbance rejection \cite{mohd2019robust}. Originally developed within the framework of variable structure systems in the mid-20th century, SMC has since been applied across diverse domains, including robotics, industrial automation, electric machinery, and aerospace systems. Its relevance to PMSM control stems from its ability to handle the nonlinear, coupled, and parameter-sensitive nature of the motor dynamics with high precision and resilience. Despite these advantages, conventional SMC suffers from a significant drawback, chattering, which manifests as high-frequency oscillations in the control signal. This phenomenon, caused by the discontinuous nature of the control input, can induce mechanical wear, excite unmodeled dynamics, and impair tracking performance \cite{chen2019precision}. To mitigate this issue, numerous enhancements to the basic SMC framework have been proposed. Most SMC designs rely on Lyapunov-based asymptotic stability analysis, often utilizing linear switching manifolds to guarantee convergence. Various Lyapunov functions have been formulated to support different system structures and sliding surface designs \cite{luo2016novel, mohd2019robust}. To further improve system performance and reduce chattering, a wide range of advanced SMC variants have been introduced in recent years. These include higher-order SMC, terminal SMC, adaptive SMC, fractional-order SMC, and integral SMC, each aiming to reduce chattering and enhance smoothness, convergence speed, and implementation feasibility while preserving the robust nature of the earlier sliding mode control.
While various studies have reviewed general PMSM control methods or SMC theory, few works provide a dedicated and up-to-date survey on the application of SMC strategies to PMSM systems. Prior surveys often lack.

\begin{itemize}
    \item Clear linkage between PMSM-specific challenges and corresponding SMC techniques,
    \item A systematic classification of modern SMC variants,
    \item In-depth discussion of implementation trends such as observer integration, optimization-aided design, and chattering mitigation.
\end{itemize}

These limitations reduce the practical utility of such reviews for engineers or researchers working directly on PMSM drives. 
With the rapid advancement in electric drive applications, ranging from intelligent transportation to renewable-powered actuators, there is an increasing demand for robust, efficient, and real-time implementable control techniques. In this context, SMC continues to evolve, yet its many variants are scattered across the literature without consolidated guidance.

This review aims to fill that gap by synthesizing the most recent developments in SMC methods specifically for PMSMs. By synthesizing recent developments and categorizing emerging strategies, we strive to provide critical insight into the effectiveness, limitations, and trade-offs of each SMC variant. Our review extends beyond mere classification to provide a comparative evaluation, implementation considerations, and potential research directions, making it a valuable resource for advancing robust control of PMSMs in uncertain and dynamic environments.

\subsection{Scope and Contributions of the Review}
\label{scope}

This review is structured to provide a focused and critical examination of SMC strategies tailored explicitly for PMSM applications. Rather than treating SMC as a generic control tool, we analyze how its diverse variants are adapted to meet the unique demands of PMSM dynamics.

The core contributions of this review are as follows:

\begin{itemize}
    \item We provide a unified taxonomy of SMC techniques used for PMSM control, covering conventional, higher-order, terminal, fractional-order, adaptive, and integral sliding modes.
    
    \item Each technique is evaluated based on its robustness properties, chattering suppression capability, convergence behavior, and suitability for implementation in practical PMSM systems.
    
    \item Recent hybrid approaches, such as those combining SMC with intelligent algorithms (e.g., PSO, HHO, fuzzy logic), observer design, and optimization frameworks, are examined in depth.
    
    \item A comparative summary table is presented to highlight the trade-offs among different SMC methods, guiding control engineers in selecting the technique based on system priorities.

    \item We provide comparative tables that summarize the strengths, limitations, and application scenarios of each method.
    
    \item Finally, we identify key research gaps and outline future research directions, particularly in the context of real-time implementation, handling of unmatched disturbances, and integration with learning-based control.
\end{itemize}

The remainder of this paper is organized as follows: Section II outlines the methodology for collecting the literature. Section III summarizes the evolution of SMC techniques for PMSM. Section IV develops a detailed taxonomy and long-table summary
of recent works. Section V presents quantitative publication trends, and Section VI discusses research challenges and open problems. The conclusions in Section VII connect these findings to the simulation study reported in Part II.

\section{Methodology}
This paper provides an in-depth analysis of sliding-mode control (SMC) techniques for the speed control of permanent magnet synchronous motors (PMSMs). An overview of basic concepts is given in \ref{sec: intro}, which is then followed by an outline of the reasons for conducting the review in \ref{scope}, as well as an overview of some selective algorithms. Based on the findings gathered from the literature, this section discusses an evaluation of key findings and outlines possible future work. The literature survey on the speed control of the PMSM was done using \textit{Scopus}, as discussed in \ref{sec:peer-review}. In this current review, the authors deliberately restrict their attention to articles published within the 2020-2025 span, aiming to capture both recent advances and the bright prospects in the field.

Chief attention was given to the identification of the papers based on the application of SMC for the speed control of PMSM. Papers were shortlisted based on their relevance to the assigned theme, ensuring they are directly relevant to the speed control of the PMSM using sliding mode.

The selection criteria were as follows:
\begin{itemize}
    \item \textbf{Relevance to PMSM speed control:} Only studies focusing on the application of the sliding mode control for speed control of the PMSMs were included.  
    \item \textbf{Publication Time frame:} It consists of the publications during the years 2020-2025, thereby making sure it covers the most recent advances in the area.
\end{itemize}

Once the relevant papers had been identified, data extraction was conducted to identify critical technical aspects, such as the type of SMC used, the observers involved, and the optimization methodologies employed. The data were then presented in a table for comparison between the studies.

The synthesis of results was conducted through:

\begin{itemize}
    \item \textbf{Categorization:} Research investigations were categorized based on the type of SMC strategy used (e.g., first-order, second-order, terminal, non-singular SMC) and the use of any observers or optimization techniques.
    \item \textbf{Comparison:} The characteristics of the selected methods were compared, focusing on the type of SMC, the observers and optimization methods employed, and the key findings reported by the authors.
    \item \textbf{Trends and Gaps:} The review highlights emerging trends in the field, such as the increasing use of intelligent optimization techniques, including reinforcement learning (RL) and neural networks (NN), as well as gaps in current research, including the real-time implementation and multi-machine control systems.
\end{itemize}

The findings of the shortlisted studies are summarized in Table~\ref{tab: smc-summary}, which presents an overview of SMC strategies, the optimization methodologies applied in each study, and the key outcomes.

\section{Structure and Characteristics of Permanent Magnet Synchronous Motors (PMSMs)}

PMSMs are widely used in various industrial and transportation applications, including electric vehicles, robotics, and renewable energy systems, due to their high efficiency, high power density, and excellent dynamic performance \cite{zhao2023review, abdelaziz2025comprehensive}. PMSMs are a class of AC motors that employ permanent magnets embedded in or attached to the rotor to create a constant magnetic field. This eliminates the need for rotor excitation current, improving efficiency and reducing losses associated with rotor windings. PMSMs are widely favored in applications requiring high-performance motion control due to their high power density, fast dynamic response, and precise torque control.

\begin{figure}
    \centering
    \includegraphics[width=0.5\linewidth]{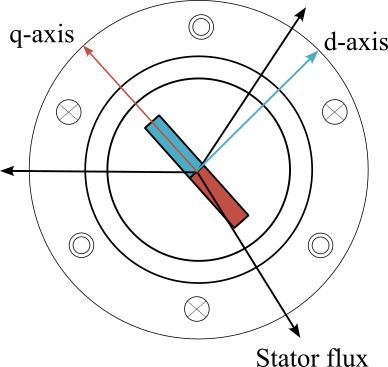}
    \caption{PMSM Internal Schematic}
    \label{fig: rotor_stator}
\end{figure}

A PMSM typically consists of a stator equipped with three-phase sinusoidally distributed windings and a rotor embedded with high-strength rare-earth permanent magnets (e.g., NdFeB) \cite{10648667,shaik2020design}. The interaction between the stator-generated rotating magnetic field and the rotor's constant magnetic field produces electromagnetic torque, resulting in synchronous rotation with the stator field (see Figure~\ref{fig: rotor_stator}). The mechanical speed of the rotor is directly proportional to the electrical supply frequency, ensuring a tight speed-frequency relationship and eliminating slip, unlike induction motors.

PMSMs are generally categorized based on the configuration of their rotor magnets \cite{abdelaziz2025comprehensive,jung2023analysis}: 

\begin{itemize}
    \item \textbf{Surface-mounted PMSM (SPMSM)}: In this type, permanent magnets are affixed to the rotor surface, resulting in a uniform air-gap flux distribution. SPMSMs typically exhibit nearly equal direct-axis and quadrature-axis inductances ($L_d \approx L_q$), leading to simple modeling and control but limited reluctance torque contribution.
    \item \textbf{Interior PMSM (IPMSM)}: Here, magnets are embedded within the rotor core. This arrangement introduces saliency in the rotor, with ($L_d \neq L_q$), enabling exploitation of reluctance torque in addition to magnet torque. IPMSMs are particularly well-suited for wide-speed-range applications and flux-weakening control.
\end{itemize}

Further classification is based on the orientation of magnetic flux lines relative to the rotor shaft \cite{eker2024experimental,nyitrai2023magnetic,zhang2023research}:

\begin{itemize}
    \item \textbf{Radial Flux PMSM}: These are the most common type, where the magnetic flux path is perpendicular to the axis of rotation. They are widely used in conventional motor drives.
    \item \textbf{Axial Flux PMSM}: In these machines, the magnetic flux flows parallel to the shaft axis. Their disc-shaped design enables high torque density and compact integration, particularly in electric vehicle drive trains and aerospace actuators.
\end{itemize}

In most practical applications, PMSMs are driven by voltage source inverters (VSIs) (see Figure~\ref{fig:pmsm_scheme}) that convert a DC supply (e.g., from a battery or rectifier) into three-phase AC voltages.  with controlled frequency and amplitude \cite{7124504}. Pulse Width Modulation (PWM) techniques are commonly used to regulate inverter output, minimizing harmonic distortion and torque ripple. Proper coordination between inverter control and PMSM dynamics is crucial for achieving the desired performance \cite{bushra2024comprehensive}.

\begin{figure}[htbp]
    \centering
    \includegraphics[width=\linewidth]{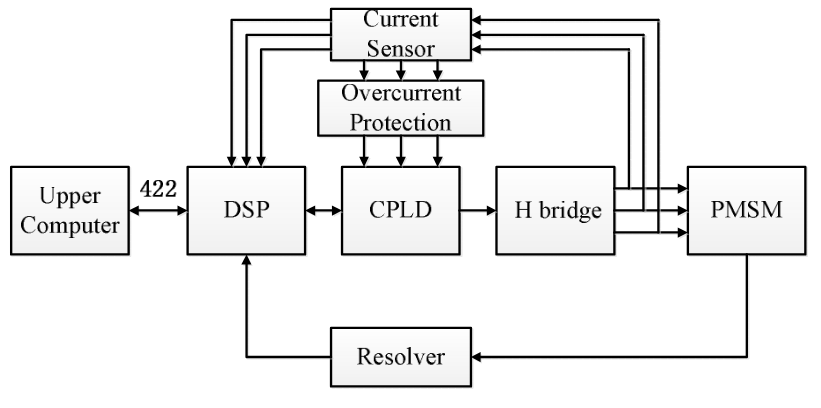}
    \caption{Block diagram of a typical PMSM control system, showing the DSP, feedback sensors, CPLD logic, inverter, and resolver-based position feedback \cite{zhou2015hardware}.}
    \label{fig: PMSMcontrl}
\end{figure}
To complement the motor’s structural characteristics, PMSMs are commonly integrated into digitally controlled drive systems. These components include a digital signal processor (DSP) for implementing control algorithms, sensors for providing current and position feedback, and an inverter stage for delivering power. Figure \ref{fig: PMSMcontrl} shows a typical block diagram of such a system, illustrating the key components involved in real-time motor control.

\section{System Model and Control Goals}
\subsection{System Model}

The mathematical modeling of a Permanent Magnet Synchronous Motor (PMSM) serves as a foundation for controller design, enabling analytical characterization of its behavior. To simplify the model while preserving the essential dynamics, the following assumptions are typically made:

\begin{itemize}
    \item The stator windings are sinusoidally distributed;
    \item Magnetic saturation, core losses, and eddy currents are neglected \cite{qian2016research};
    \item The motor is symmetric and balanced;
    \item The system is transformed to the synchronous $d$-$q$ reference frame via Park’s transformation \cite{abraham2024speed,peng2023overview}
\end{itemize}

Under these assumptions, the PMSM voltage equations in the rotating $d$-$q$ reference frame are given by

\begin{align}
\frac{d i_d}{dt} &= \frac{1}{L_d} \left( u_d - R_s i_d + \omega_e L_q i_q \right) \\
\frac{d i_q}{dt} &= \frac{1}{L_q} \left( u_q - R_s i_q - \omega_e (L_d i_d + \psi_f) \right)
\end{align}

Where:
\begin{itemize}
    \item $i_d$, $i_q$ are the direct and quadrature axis stator currents,
    \item $u_d$, $u_q$ are the corresponding stator voltages,
    \item $L_d$, $L_q$ are the stator inductances,
    \item $R_s$ is the stator resistance,
    \item $\omega_e$ is the electrical angular velocity,
    \item $\psi_f$ is the rotor flux linkage.
\end{itemize}

The electromagnetic torque $T_e$ produced by the PMSM is defined as
\begin{equation}
T_e = \frac{3}{2} p_n \left[ \psi_f i_q + (L_d - L_q) i_d i_q \right]
\end{equation}

In surface-mounted PMSMs, where $L_d = L_q$, the reluctance torque term vanishes, simplifying the torque equation to:

\begin{equation}
T_e = \frac{3}{2} p_n \psi_f i_q
\end{equation}

Newton's second law governs the mechanical dynamics of the rotor:

\begin{equation}
J \frac{d \omega_r}{dt} = T_e - B \omega_r - T_L
\end{equation}

Where:
\begin{itemize}
    \item $J$ is the moment of inertia,
    \item $\omega_r$ is the rotor speed,
    \item $B$ is the viscous friction coefficient,
    \item $T_L$ is the load torque.
\end{itemize}

The relationship between electrical and mechanical angular speeds is
\begin{equation}
\omega_e = p_n \omega_r
\end{equation}

Similar modeling is also described in \cite{wang2024disturbance} and \cite{zhang2024improved}.

\subsection{Control Objectives}

The primary control goals for PMSM systems are

\begin{itemize}
    \item \textbf{Accurate Speed Regulation:} Maintain rotor speed $\omega_r$ at a desired setpoint $\omega^*$, despite variations in load torque $T_L$ or system parameters.
    \item \textbf{Efficient Torque Control:} Ensure precise tracking of desired torque $T_e^*$ to meet dynamic response specifications.
    \item \textbf{Robustness:} Achieve control objectives under parametric uncertainties (e.g., changes in $R_s$, $J$, or $\psi_f$) and external disturbances.
    \item \textbf{Reduced Chattering:} For robust control techniques like SMC, minimizing high-frequency control oscillations is essential for practical deployment.
\end{itemize}

\begin{figure}
    \centering
    \includegraphics[width=\linewidth]{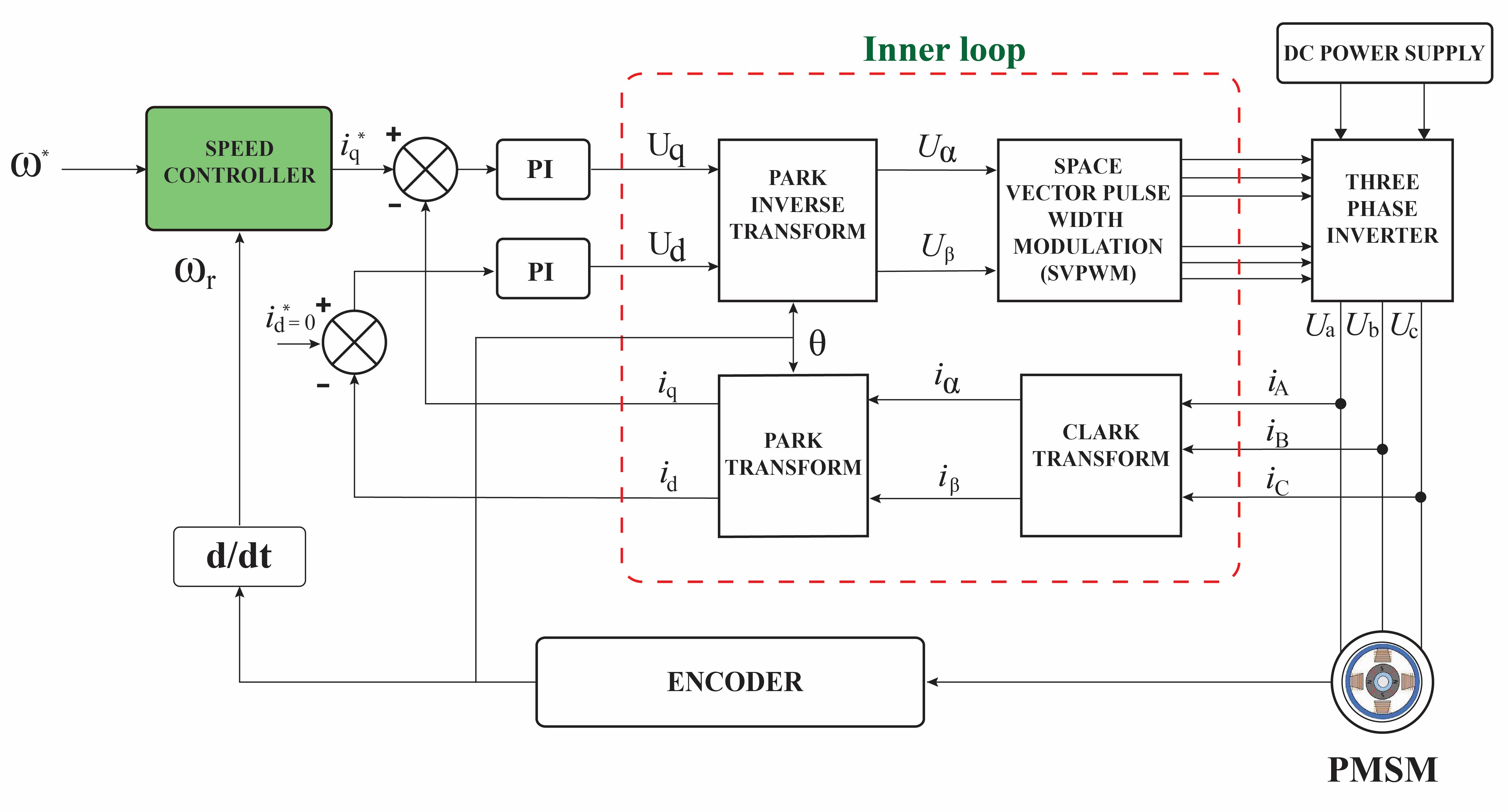}
    \caption{Complete control architecture of a PMSM drive based on FOC. The system comprises an outer speed control loop, inner current regulation loops, Park and Clarke transformations, SVPWM, and encoder-based feedback.}
    \label{fig:pmsm_scheme}
\end{figure}

These objectives serve as the foundation for the various SMC-based control strategies presented in the following sections. Figure~\ref{fig:pmsm_scheme} illustrates a typical field-oriented control (FOC) scheme implemented in PMSM drives. It includes cascaded speed and current controllers, coordinate transformations, space vector pulse-width modulation (SVPWM), and encoder feedback. This architecture serves as the foundation upon which more advanced control methods, including SMC, can be integrated.

\section{Sliding Mode Control (SMC) for PMSM}
SMC is a nonlinear control technique renowned for its robustness, finite-time convergence, and strong rejection properties against disturbances \cite{ullah2022critical, 10990020}. It originated from the theory of variable structure control systems in the mid-20th century and has since been widely applied in motion control, robotics, power converters, and electromechanical systems \cite{gambhire2021review,utkin1977variable}. Its key strength lies in its ability to drive system trajectories onto a predefined sliding surface and maintain them there, ensuring predictable and resilient closed-loop dynamics \cite{hosseyni2018new}.

In the context of PMSM drives, SMC has gained significant traction due to its inherent suitability for handling the nonlinearities, parameter uncertainties, and load disturbances commonly encountered in high-performance motor applications. Conventional controllers such as PI or model-based FOC schemes often struggle under these conditions, particularly when system parameters drift due to temperature variations, load torque changes, or sensor noise. SMC offers an appealing alternative by decoupling the system's response from such variations once the system is in the sliding regime.

The basic structure of SMC involves two key components: \begin{itemize}
    \item Designing a sliding surface that encapsulates the desired system dynamics, and
    \item Synthesizing a discontinuous control law that forces the system states to reach and remain on this surface. For example, in PMSM speed control applications, the sliding surface is often constructed using tracking errors in speed and/or current \cite{hosseyni2018new}.
\end{itemize} 

\begin{equation} \label{slid1}
s(t) = c e(t) + \dot{e}(t)
\end{equation}

where $e(t) = \omega^* - \omega_r$ is the speed tracking error, $\dot{e}(t)$ its derivative, and $c > 0$ is a design constant satisfying the Hurwitz condition. The control law is then derived to ensure the sliding condition:

\begin{equation}
\dot{V} = \frac{1}{2} \frac{d}{dt}s^2(t) < 0, \quad \text{for } s(t) \neq 0
\end{equation}

\begin{figure}
    \centering
    \includegraphics[width=0.5\linewidth]{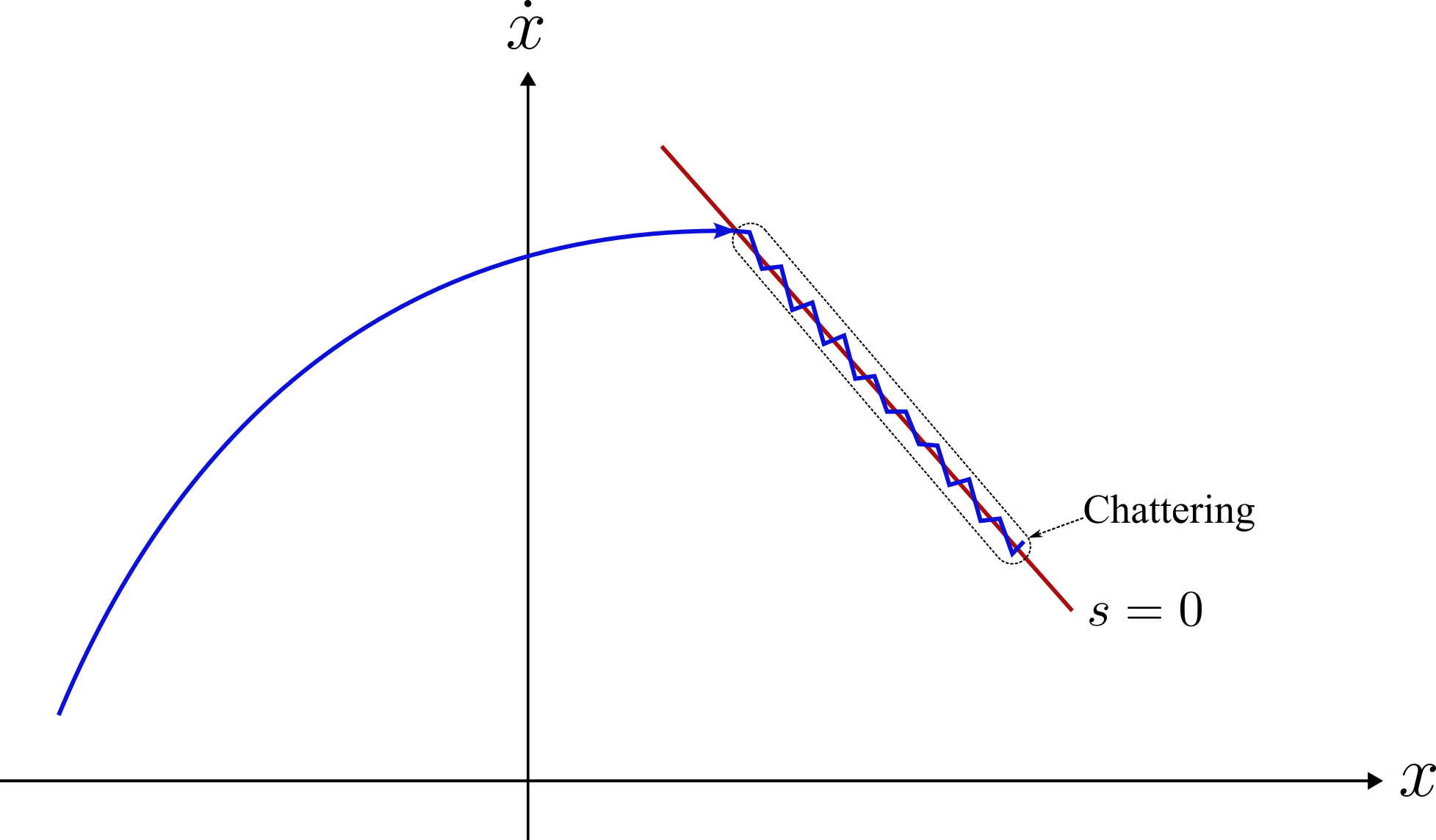}
    \caption{Chattering in Conventional SMC}
    \label{fig: chattering}
\end{figure}

While conventional SMC ensures robust and accurate tracking, its discontinuous nature leads to a significant challenge known as \textit{chattering}, high-frequency switching in the control signal (see Figure \ref{fig: chattering}). This phenomenon can excite unmodeled dynamics, cause mechanical wear, and generate electromagnetic interference, especially in sensitive motor drive systems \cite{chen2019precision, 9645927, saif2023fractional}. To address these drawbacks, numerous advancements have been proposed. These include HOSMC, which smoothens the control by acting on higher-order derivatives; terminal and fast terminal SMC, which provide finite-time convergence; and adaptive or fractional-order SMC techniques, which enhance robustness and flexibility. Many of these approaches incorporate continuous approximations of the sign function (e.g., saturation or tanh functions) and use observer-based feedback to reduce sensor noise effects.

In PMSM applications, SMC has been employed for speed control, torque regulation, current tracking, and sensorless operation. It has also been integrated with optimization algorithms (e.g., PSO, HHO), intelligent observers (e.g., ESO, DOB), and model predictive control (MPC) to enhance performance and adaptability under real-time constraints. The following sections present a structured review of state-of-the-art SMC techniques tailored to PMSM systems. Each variant is discussed in terms of its theoretical foundation, implementation considerations, and demonstrated performance in PMSM-related literature.

\begin{figure*}
    \centering
    \includegraphics[width=\textwidth]{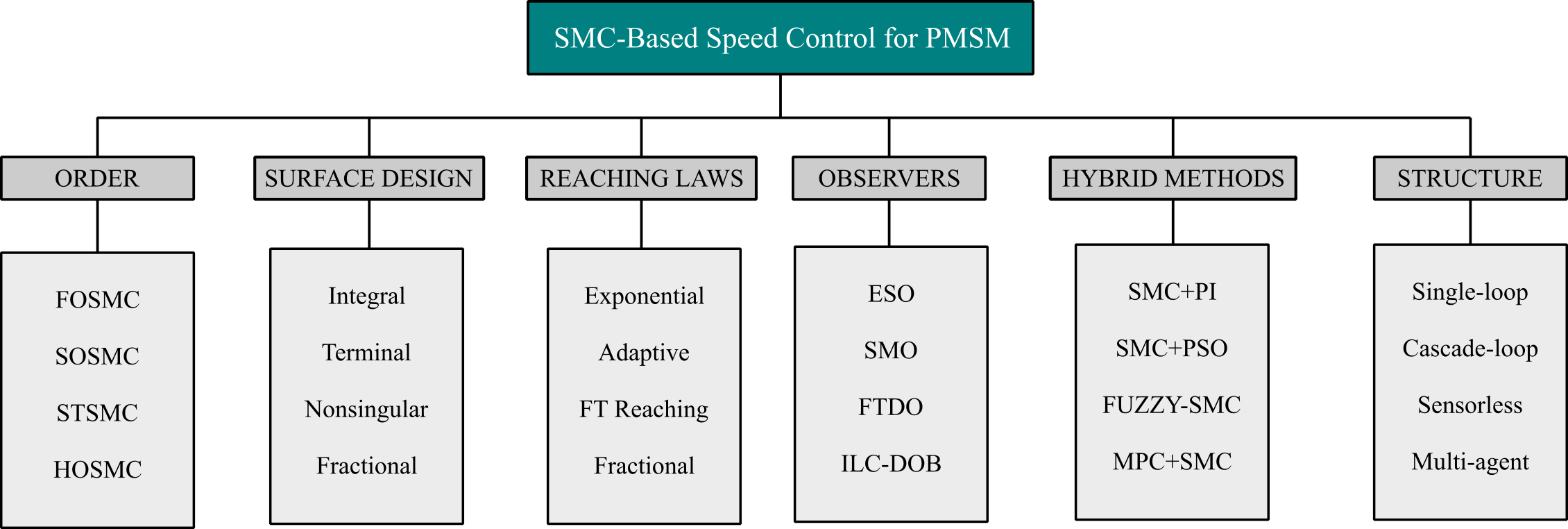}
    \caption{Taxonomy of Sliding Mode Control (SMC) strategies for PMSM speed regulation, covering control orders, surface designs, reaching laws, observers, hybrid methods, and structural configurations.}
    \label{fig: smc_taxonomy}
\end{figure*}

To provide a comprehensive overview of the design landscape, Figure~\ref{fig: smc_taxonomy} presents a taxonomy of SMC-based strategies tailored for PMSM speed regulation. The classification highlights key control features such as control order (e.g., FOSMC, STSMC), surface design (e.g., integral, terminal, non-singular), reaching laws (e.g., adaptive, finite-time, fractional), observer types (e.g., ESO, SMO, FTDO), hybrid control combinations (e.g., SMC+PSO, FUZZY-SMC), and system structures (e.g., single-loop, cascade, sensorless). This diagram serves as a high-level map to guide the detailed discussion in the subsequent sections.

\subsection{Conventional Sliding Mode Control}
The conventional SMC is also known as the traditional, or first-order SMC. It represents the classical implementation of the sliding mode framework, characterized by the use of a linear switching surface and a discontinuous control input \cite{utkin1977variable}. Its design methodology revolves around two primary steps: (i) constructing an appropriate sliding surface that encodes the desired system behavior and (ii) synthesizing a control law that guarantees the finite-time convergence of system trajectories to the sliding surface and maintains motion along it thereafter \cite{li2024overview}. For PMSM speed control, a common sliding surface is defined in terms of the tracking error as shown in \eqref{slid1}: 

where $e(t) = \omega^* - \omega_r$ is the speed tracking error, $\dot{e}(t)$ is its time derivative, and $c > 0$ is a positive design parameter satisfying the Hurwitz condition. The control objective is to ensure $s(t) \rightarrow 0$ in finite time and remain on the sliding manifold $s(t) = 0$ thereafter.

To achieve this, a Lyapunov function is typically chosen as:

\begin{equation} \label{lf1}
V(t) = \frac{1}{2} s^2(t)
\end{equation}

And its time derivative is enforced to be negative definite:

\begin{equation}
\dot{V}(t) = s(t)\dot{s}(t) < 0, \quad \forall s(t) \neq 0
\end{equation}

A commonly used control law that satisfies this condition is the sign-based switching control:

\begin{equation}
u(t) = u_{eq}(t) - k\, \text{sgn}(s(t))
\label{eq:smc_control}
\end{equation}

Where $u_{eq}(t)$ is the equivalent control (typically derived via model dynamics), $k > 0$ is the switching gain, and $\operatorname{sgn}(\cdot)$ is the signum function. This control drives the system state toward the sliding surface and compensates for matched disturbances and parameter variations. Based on \eqref{lf1}, the sliding surface is designed such that the following conditions are met:

\begin{align}
\begin{cases} 
\dot{V} \leq 0, \\
V > 0, \ \forall S \neq 0, \\
V = 0, \ \text{when } S = 0, \\
\dot{V} < 0, \ \forall S \neq 0.
\end{cases}
\end{align}

While conventional SMC offers robustness and simplicity, it is prone to a critical drawback known as \textit{chattering}. Chattering refers to high-frequency oscillations in the control input due to the discontinuous switching action \cite{liu2024non}. This effect can excite unmodeled system dynamics, cause mechanical wear in actuators, and degrade control performance, especially in high-precision PMSM applications. To alleviate these issues, several techniques have been explored, including the use of boundary layers (e.g., replacing $\text{sgn}(s)$ with $\tanh(s/\epsilon)$ or saturation functions) and, more importantly, the development of higher-order and continuous SMC variants. These approaches aim to retain the robustness of conventional SMC while significantly reducing or eliminating chattering.

Several refinements of conventional SMC have been proposed in the literature to improve speed regulation performance and mitigate chattering in PMSM drives, such as integration with reinforcement learning \cite{nicola2022improvement}. In \cite{wang2019new}, an innovative sliding mode controller (SMC) incorporating a modified reaching law was integrated with an extended state observer (ESO) to estimate cumulative disturbances, providing feedforward correction to enable robust tracking with minimal chattering. In a different investigation, the authors in \cite{feng2020speed} addressed disturbance and parameter variation by proposing a sliding mode control strategy that incorporates state vectors into the reaching law, thereby enhancing dynamic response and overall system robustness under varying operating conditions. In \cite{yeam2020design}, a sliding mode controller was developed for dual PMSMs with a single inverter, applying vector control to the master and damping control to suppress load-induced speed oscillations. The approach demonstrated robust performance under varying parameters. Other methods include augmenting SMC with equivalent input disturbance control (EIDC), which has been shown to attenuate disturbances effectively and suppress chattering under direct torque control (DTC) frameworks \cite{fang2022speed}. Sliding mode technique was employed in \cite{hou2022sliding} to address the challenge of system uncertainties in vector control; the authors proposed a cascaded variable-rate SMC with an exponential reaching law within a sliding mode predictive current control framework, which effectively enhances system A non-cascade sliding mode control strategy for surface-mounted PMSMs was proposed in \cite{che2022singular}, it uses singular perturbation theory and a nonlinear tracking differentiator to improve system stability and robustness while reducing chattering and mitigating control signal jitter. To address the discontinuity of the sign function, several approaches, such as modifying the reaching law \cite{usama2021low, zhang2024novel} and saturation-based modifications, have been proposed. For instance, \cite{peng2022novel} implemented a vector control strategy incorporating a novel sliding mode approach law where the $\text{sgn}(\cdot)$ function was replaced with a continuous saturation function, leading to improved smoothness and performance. Similarly, bio-inspired optimization has also been utilized to tune SMC parameters. To improve PMSM speed control, a backstepping sliding mode controller with a nonlinear disturbance observer (NDO-SMC) was suggested in \cite{duan2022backstepping}, which, when compared to regular SMC and I-SMC, reduces overshoot, settling time, and torque ripple. Authors in  \cite{wei2022improved} also suggested an enhanced sliding mode control technique with a saturation function. The method improved starting performance and speed stability by combining PI control and controlling the inverter switching frequency, which decreased speed fluctuation from 7.7 \% to 5.9 \%. In \cite{choi2023sliding}, a whale optimization algorithm was applied to design a sensorless SMC scheme, yielding high reliability and resilience. Likewise, Harris Hawk Optimization (HHO) was utilized in \cite{abraham2024speed} to minimize speed fluctuations and enhance robustness against load disturbances. For further tuning flexibility, \cite{yuan2024novel} proposed a simulated annealing particle swarm optimization (SAPSO) technique with an ESO and nonlinear state error feedback (NLSEF) applied to PMLSMs.

Several studies have also emphasized the integration of disturbance observers with SMC \cite{qu2021design}.
In a further contribution in \cite{ma2020active}, an active disturbance compensation strategy was proposed for PMSM speed control, using SMC–ESO in the speed loop and ESO-based PI control in the q-axis loop. Simulations and experiments confirmed improved disturbance rejection and responsiveness. To eliminate chattering and speed up reaching time, a composite SMC that combines a hybrid reaching law (HRL) and an extended sliding mode disturbance observer (ESMDO) was suggested in \cite{sun2021composite}. The performance of the proposed controller was experimentally validated and compared with the conventional modified reaching law under various operating conditions. Authors in \cite{wang2021antidisturbance} designed an improved Deadbeat DTC (DBDTC) based on SMC, implementing an antidisturbance SMC that incorporates an exponential reaching law along with an extended sliding mode observer to address the suboptimal precision and poor dynamic behavior of the conventional DBDTC; this approach effectively improves convergence while also suppressing control jitter. In \cite{gil2023nonlinear}, a nonlinear SMC (NSMC) with a disturbance observer (DOB) outperformed conventional SMC in suppressing chattering via a nonlinear gain function on the sliding surface. An augmented reaching law and observer-enhanced SMC was introduced in \cite{guo2022fast, chen2024variable}, resulting in faster convergence and enhanced disturbance rejection. Similarly, \cite{hong2024composite} proposed a composite SMC scheme that combines sliding mode observers and first-order differentiators, which was validated through experimental results. Alternative reaching laws have also been explored for smoother control action. For example, \cite{gong2022robust} proposed an improved power reaching law (IPRL) utilizing a hyperbolic tangent function instead of discontinuous control logic. A modified reaching law with an exponential sliding surface was introduced in \cite{chiliveri2023modified} to handle load torque disturbances and parameter mismatches. In \cite{qu2023sliding}, an advanced sliding mode reaching law (ASMRL) was employed, combining power terms and saturation functions to enhance convergence and robustness. Similarly, a hybrid reaching law consisting of a finite-time and exponential term 
was incorporated with load observers in \cite{zhang2023time} to reduce chattering and improve system performance; experimental results showed better performance under different operating conditions.  Iron-loss-aware control design using reaching law techniques was presented in \cite{huang2024novel} to achieve high dynamic performance and efficiency simultaneously.

Recent improvements in embedded hardware and processing capability have further enabled hybrid SMC structures. In \cite{zhou2023novel}, an optimized proportional resonant (SMC-PR) controller was integrated with SMC, showing improved dynamic tracking on the MATLAB/Simulink platform. \cite{yang2025combined} also employed a hybrid approach, which incorporates SMC and state feedback to mitigate modeling disturbances in PMSMs. A fixed-time convergent SMC and observer pair (FSMC/FSMO) was proposed in \cite{zhang2024fixed} to ensure convergence within a finite time horizon. Additionally, a barrier-function-based SMC (BFSMC) with DOB was introduced in \cite{dai2024disturbance} to tackle speed fluctuation constraints and maintain control continuity. Incorporating predictive elements, \cite{zhou2024sliding} developed a model predictive current control (MPCC) scheme with a sliding mode disturbance observer and load torque estimation. Furthermore, \cite{zhang2024enhanced} proposed an adaptive sliding mode reaching law (ASMRL) to address slow convergence and residual chattering observed in standard SMC. A fuzzy controller topology was introduced in \cite{qiao2024fuzzy} to improve robustness in the presence of system fluctuation.

Conventional SMC, despite its limitations, remains a foundational method from which numerous modern control techniques for PMSM have evolved. The cited works illustrate a diverse range of enhancements designed to retain the robustness of SMC while improving its smoothness, adaptability, and practicality of implementation. The following sections provide a detailed overview of further advancements.

\subsection{Terminal Sliding Mode Control (TSMC)}

Terminal Sliding Mode Control (TSMC) is an advanced variant of the conventional Sliding Mode Control (SMC) designed to achieve finite-time convergence of system states to the equilibrium point, in contrast to the asymptotic convergence offered by conventional SMC \cite{gambhire2021review,venkatarman1992control}. This characteristic makes TSMC especially attractive for high-performance applications such as PMSM speed control, where fast and accurate tracking is essential \cite{mu2017dynamic}.

The key innovation in TSMC lies in its non-linear sliding surface formulation. A typical terminal sliding surface is expressed as

\begin{equation}
s(t) = \dot{e}(t) + \beta |e(t)|^{\lambda} \text{sgn}(e(t))
\label{eq:tsmc_surface}
\end{equation}

where $e(t)$ is the tracking error, $\beta > 0$ is a positive constant, and $0 < \lambda < 1$ defines the nonlinearity degree. This design ensures that the convergence rate accelerates as the error diminishes, resulting in finite-time stabilization. The control law is then derived to enforce $\dot{V} = \frac{1}{2} \frac{d}{dt}s^2(t) < 0$, ensuring stability in the Lyapunov sense. While TSMC offers enhanced convergence and robustness compared to conventional SMC, early formulations were affected by singularity issues when the error $e(t)$ approached zero. To resolve this, various non-singular terminal sliding mode control (NTSMC) \cite{wang2024single, chen2024non} and fast-terminal sliding mode control (NFTSMC) techniques were introduced, modifying the sliding surface to avoid division by zero or undefined dynamics near the origin.

Terminal Sliding Mode Control (TSMC) has been extensively explored to enhance the performance of the PMSM systems, particularly in terms of speed \cite{zhou2025adaptive, hu2023speed} and position \cite{che2024barrier} regulation. In \cite{zhu2020position}, a neural network (NN)-based terminal sliding mode control strategy was employed in a PMSM servo system, demonstrating rapid convergence and strong disturbance rejection. Neural networks were combined with fractional-order complementary non-singular TSMC in \cite{zhang2024fractional}, aiming to address uncertainties. This method combines generalized and complementary sliding mode surfaces, utilizing fractional calculus operators to improve position tracking accuracy and reduce chattering. The neural network component estimates system uncertainties in real-time, thereby enhancing dynamic response and anti-interference capabilities. Another significant development is the fuzzy Global Fast TSMC (GFTSMC) strategy, which incorporates a Luenberger disturbance observer. This approach employs fuzzy logic (FL) to adjust controller parameters in real-time, effectively suppressing chattering without compromising convergence speed. The Luenberger observer estimates disturbances, feeding back the values to the controller to mitigate their impact on the PMSM system \cite{xbzrb2023fuzzygftsmc}. 

To address communication constraints in networked PMSM systems, event-triggered versions of TSMC have been proposed. This method integrates a Genetic Algorithm (GA)-optimized ESO to estimate perturbations, reducing chattering in the control signal. The periodic event-triggered mechanism schedules signal transmission efficiently, thereby decreasing communication overhead while maintaining system stability \cite{song2022periodic,yukun2022periodic}. Building on this, an adaptive TSMC approach utilizing an NN-based disturbance estimation has been introduced. This dynamic event-triggered strategy addresses the need for real-time adaptation to disturbances and parameter uncertainties, ensuring robust speed regulation in PMSM systems \cite{song2023adaptive}.

Control ripple suppression and rapid convergence to the sliding surface were optimized in \cite{junejo2022novel} via a fast terminal reaching law with dual power expressions, integrated with ESO to provide feedforward correction. The designed framework proved effective both computationally and experimentally, significantly improving overall control performance. Similarly, \cite{zhang2023pmsm} presented an NFTSMC design incorporating a novel reaching law alongside a disturbance observer (NFTSMDO), which yields improved speed and torque tracking performance, as verified through simulations using MATLAB/Simulink and dSPACE. Observer-based enhancements have also been explored to enhance robustness and estimation accuracy. In \cite{li2022adrc}, an NFTSM observer was integrated into an Active Disturbance Rejection Control (ADRC) framework to achieve accurate estimation of position and velocity states in PMLSMs. Furthermore, to enhance the control performance of surface-mounted PMSM (SPMSM) systems, an Improved Non-Singular Fast Terminal Sliding Mode Control (INFTSMC) strategy was proposed in \cite{kim2023improved}. By eliminating singularities in the controller input, the approach demonstrated high tracking precision, rapid convergence, and strong resilience to disturbances.

TSMC has also proven effective in fault-tolerant scenarios. In \cite{ma2022fault}, an NTSMC-based control was implemented in a three-phase, four-switch PMSM drive to ensure seamless switching and reference tracking during open-circuit faults. Similarly, \cite{wang2023second} combined NFTSMC with a robust compensator and velocity disturbance observer to suppress interference and improve tracking precision in robotic PMLSM systems. To further reduce chattering and improve error convergence, high-order extensions and composite designs have been developed. For example, a second-order nested TSMC (SNTSMC) for PMSM current control was proposed in \cite{dongye2023second}, which embedded integral fast terminal sliding dynamics within a nested sliding structure. The approach achieved rapid convergence and significantly minimized steady-state error. Similarly, \cite{le2023speed} proposed a high-order TSMC (TSMHC) for PMSM drives under FOC, combining terminal and higher-order elements to improve stability under load disturbances.

Terminal controllers have also been significantly utilized with observers \cite{hou2024fixed, wang2024speed, tian2024integrated} to mitigate the effect of chattering. A sliding mode extended state observer (SMESO) built upon FTSMC was proposed in \cite{xu2020efficient}. The proposed controller demonstrated reduced chattering, while also improving disturbance rejection with minimal steady-state error. In relation to this, an adaptive terminal sliding mode reaching law (ATSMRL) was integrated with a continuous fast terminal sliding mode control (CFTSMC) in \cite{junejo2020adaptive}, enabling accurate trajectory tracking while incorporating an extended sliding mode disturbance observer (ESMDO) to enhance disturbance rejection. In a similar vein, a fast terminal sliding mode control (FTSMC) built on a finite time sliding mode observer (FTSMO) was proposed in \cite{xu2021composite}. The proposed framework was demonstrated to exhibit resilience against interference, robust trajectory tracking, and continuous system performance in different operating regimes. In a different study, a backstepping-based NTSMC with a bounded-time disturbance observer was proposed in \cite{li2021backstepping}, where the controller demonstrated asymptotic steadiness with finite-time convergence and strong resilience to parameter uncertainties.

In a separate study, a hybrid approach incorporating fixed-time observers and non-singular terminal control was proposed in \cite{wu2023hierarchical} to reduce tracking error while minimizing chattering. Super-twisting observers \cite{chen2023continuous} were incorporated to enhance speed convergence, while the robustness of the proposed controllers was demonstrated via experimentation. Optimization-based disturbance observers with feedforward compensation were utilized with fast integral terminal controllers in \cite{yang2023fast}. This combination provided remarkable interference rejection capabilities, thus significantly reducing torque fluctuations. The superiority of the proposed method was demonstrated through comparisons with traditional approaches.
. \cite{kong2024new} utilizes a terminal control approach and observers to mitigate the effect of modeling errors. Finally, \cite{wang2023high} introduced a fast recursive TSMC (FRTSMC) with a mover-velocity disturbance observer (MVDO) to enhance anti-disturbance performance in permanent magnet linear synchronous motors. Comparative results demonstrated its superiority over standard NFTSMC and other observer-based methods in terms of error reduction and robustness. These studies confirm that TSMC and its variants, including non-singular, fast terminal, fractional-order, fuzzy, adaptive, event-triggered, and high-order structures, have emerged as robust solutions for PMSM speed and current control. Their modularity and resilience to nonlinearity, parameter drift, and external disturbances make them especially suitable for high-precision and safety-critical motor drive systems.

\subsection{Integral Sliding Mode Control (ISMC)}

While SMC offers strong robustness properties for nonlinear systems, it suffers from a significant limitation during the reaching phase: the system is not yet on the sliding surface and therefore not robust to matched disturbances. ISMC was developed to address this limitation by ensuring that the system trajectory begins on the sliding manifold from the initial time, thereby eliminating the reaching phase and guaranteeing robustness from the outset \cite{fridman2011sliding, pan2017integral,abbasi2022robust}.

The key objectives of ISMC include
\begin{itemize}
    \item Maintaining the original system's order without reducing state-space dimensionality.
    \item Ensuring robustness against matched perturbations throughout the system's evolution.
\end{itemize}

ISMC is generally formulated for a perturbed nonlinear system of the form:
\begin{equation}
\dot{x} = f(x,t) + B(x)\big(u + \delta\big) + \varphi_{\text{um}},
\label{eq:ismc_perturbed}
\end{equation}
where \(\delta\) is a matched disturbance and \(\varphi_{\text{um}}\) denotes unmatched disturbances. The ISMC strategy defines the control input as:
\begin{equation}
u = u_0(x,t) + u_1(x),
\end{equation}
Where \(u_0(x,t)\) is a nominal control input, and \(u_1(x)\) is a discontinuous term responsible for maintaining the invariance of the sliding variable with respect to \(\delta\). The integral sliding surface is given by:
\begin{equation}
s(x,t) = g(x) - z(t),
\end{equation}
\begin{equation}
z(t) = g(x_0) + \int_{t_0}^{t} G(x_0)\left[f(x_0,\tau) + B(x_0)u_0(x_0,\tau)\right] d\tau,
\end{equation}
where \(g(x)\) is a smooth function and \(G(x) = \frac{\partial g(x)}{\partial x} B(x)\) is assumed invertible. This structure ensures that~\(s(x_0, t_0) = 0\), meaning the trajectory starts directly on the sliding manifold.

ISMC has been widely applied in PMSM control to improve tracking performance and disturbance rejection. To enhance robustness, observer-based techniques have been integrated into ISMC structures. For instance, an extended state observer (ESO) was employed in \cite{song2017integral} to provide feedforward compensation and improve speed tracking. Similarly, a disturbance observer was introduced in \cite{suleimenov2019integral} for surface-mounted permanent magnet synchronous motor (PMSM) systems and validated through MATLAB/Simulink simulations. In another study, a high-order observer (HOO) was used in \cite{liu2021high} to decouple the speed and current control loops of a permanent magnet linear synchronous motor (PMLSM), achieving improved estimation of parameter uncertainties and external disturbances. To enable better transient and improved system performance, authors in \cite{lu2021disturbance} proposed a perturbation rejection approach that incorporates a resilient controller consisting of integral sliding mode control with composite nonlinear feedback (CNF); the controller also integrates a fuzzy-based sliding mode observer (SMO) to compensate for inherent chattering. Structural refinements to ISMC have also been investigated to reduce chattering and improve convergence speed. In \cite{baik2000robust}, a boundary layer ISMC strategy was developed based on a quasi-linear decoupled PMSM model, incorporating a PD controller in the outer loop. Simulation results showed that this method outperformed conventional feedback linearization in terms of speed regulation. For velocity control of PMSMs, \cite{qi2015permanent} suggests a second-order integral sliding mode control (SOISMC) algorithm that improves and guarantees system convergence using a Lyapunov function by lowering static error. To address the windup issue, the system incorporates anti-windup management in addition to eliminating chattering. Composite and fractional-order extensions of ISMC have also been explored. In \cite{huang2014composite}, a composite nonlinear sliding surface was designed by combining integer- and fractional-order integral terms. The fractional component served as feedforward compensation, enhancing robustness to load disturbances while maintaining fast response. Additionally, a model-free ISMC approach using a fast-reaching law was introduced in \cite{li2021model} for ultra-local PMSM modeling in automotive applications. The proposed controller was validated using Lyapunov stability theory and experimental testing, demonstrating strong anti-disturbance characteristics and a rapid convergence speed.

Recent advances in optimization-based tuning of ISMC have also yielded promising results. In \cite{khorsand2017optimal}, particle swarm optimization (PSO) was used to fine-tune controller parameters based on a cost function that accounted for system performance and robustness. The optimized ISMC outperformed conventional PI controllers, particularly under saturation, parameter drift, and unmodeled disturbances. To further strengthen robustness, hybrid control strategies that combine ISMC with other robust techniques have been proposed. For example, \cite{8823533} developed a dual-mode controller that integrates ISMC with \(H_\infty\) control for SPMSM systems. The combined approach showed improved rejection of both matched and mismatched disturbances, enhancing overall control precision and stability compared to standalone methods.

In summary, ISMC represents a significant advancement in robust control for PMSM systems by eliminating the vulnerability of the reaching phase. Its flexibility in integrating observers, fractional-order dynamics, optimization routines, and hybrid robust designs makes it a versatile tool for high-performance motor drive applications subject to uncertainties and external disturbances.

\subsection{Higher Order Sliding Mode Control (HOSMC)}

Higher Order Sliding Mode Control (HOSMC) has emerged as a pivotal advancement in sliding mode control theory, primarily aimed at addressing the well-known limitations of conventional first-order SMC, most notably, the chattering phenomenon arising from discontinuous control actions \cite{msaddek2014comparative}. In conventional SMC, the control input abruptly switches to enforce the sliding condition, which often leads to high-frequency oscillations that are undesirable in practical systems, such as PMSMs. HOSMC overcomes this limitation by designing control laws that not only drive the sliding surface \( s(t) \) to zero in finite time but also guarantee the convergence of its derivatives up to order \( r-1 \), where \( r \) is the system’s relative degree \cite{levant2003higher,laghrouche2007higher,li2024overview}. HOSMC targets systems with relative degree \( r > 1 \), designing control laws that enforce 
\begin{equation}
s(t) = \dot{s}(t) = \dots = s^{(r-1)}(t) = 0, \quad \text{in finite time}
\end{equation}
This results in improved control accuracy, reduced chattering, and enhanced robustness, especially in high-precision applications. 

The general formulation of HOSMC begins with a nonlinear system subject to matched uncertainties:
\begin{equation}
\label{eq:hosmc_dynamics}
\dot{x} = f(x) + \sum_{i=1}^{m} g_i(x) u_i, \qquad y = \sigma(x),
\end{equation}
where \(x \in \mathbb{R}^n\) is the state vector, \(u \in \mathbb{R}^m\) is the control input, and \(\sigma(x)\in\mathbb{R}^m\) is a smooth output function. The control objective is to drive the output \(\sigma(x)\) and its successive time derivatives to zero in finite time. Achieving this requires the following assumptions, as described in \cite{defoort2012higher}:

\begin{itemize}
  \item \textbf{Assumption 1:} The relative degree vector \(r = [r_1, \dots, r_m]^T\) with respect to \(\sigma(x)\) is known and constant. The decoupling matrix composed of Lie derivatives is non-singular, and the zero dynamics are assumed to be asymptotically stable.
  
  \item \textbf{Assumption 2:} The system is interpreted in the Filippov sense to handle discontinuities in the control signal and guarantee the existence of solutions for discontinuous vector fields.
  
  \item \textbf{Assumption 3:} The nonlinear functions \(f(x)\) and \(g_i(x)\) are decomposable into nominal parts and bounded uncertainties. The uncertainties are norm-bounded and measurable, i.e., they are contained within a known function \(\rho(x)\) with a known lower bound \(0 < \alpha \leq 1\).
\end{itemize}

Under these conditions, the HOSMC control strategy typically consists of a combination of a nominal controller and a discontinuous component designed to ensure finite-time convergence. The control law takes the form:
\begin{equation}
w = w_{\text{nom}}(z) + w_{\text{disc}}(z, z_{\text{aux}}), \qquad \dot{z}_{\text{aux}} = -w_{\text{nom}}(z),
\end{equation}
Where \(z\) represents the augmented system state and \(w_{\text{disc}} = -G(z)\,\text{sign}(s(z))\) serves as the corrective term driving the system to the higher-order sliding manifold. The function \(s(z)\) denotes the generalized sliding variable, incorporating auxiliary states that capture derivative information.

To ensure robustness and finite-time convergence, the switching gain \(G(z)\) is designed to satisfy the inequality:
\begin{equation}
G(z) \geq \frac{(1 - \alpha) \|w_{\text{nom}}(z)\| + \rho(x) + \eta}{\alpha}, \qquad \eta > 0.
\end{equation}

This formulation allows the system to reject bounded matched uncertainties without introducing excessive control activity. The higher-order convergence conditions make HOSMC especially attractive for precision applications such as PMSM control, where actuator wear, mechanical resonance, and sensitivity to measurement noise are essential considerations. By ensuring smoothness in the control signal while retaining robustness, HOSMC provides a powerful solution space that encompasses a variety of control strategies, including the widely used super-twisting algorithm, adaptive sliding strategies, and terminal HOSMC variants \cite{hamida2013high, HAMIDA201479}. 

Building upon the foundational concepts of Higher Order Sliding Mode Control (HOSMC), this section explores its practical applications and advancements in Permanent Magnet Synchronous Motor (PMSM) systems. The focus is on various HOSMC strategies, including the STA, adaptive methods, observer-based designs, and hybrid control schemes, all aimed at enhancing control performance, robustness, and reducing chattering effects.

\subsubsection{Super-Twisting Algorithm (STA) and Its Enhancements}

STA is a widely adopted second-order sliding mode technique known for ensuring finite-time convergence of the sliding variable and its derivative, thereby producing continuous control signals that alleviate chattering. Its application in PMSM systems has demonstrated improved current regulation, speed tracking, and robustness against matched disturbances. In \cite{zhang2021fast,song2024pmsm}, a fast-super-twisting SMC (FSTA-SMC) was implemented in the PMSM speed loop. The results showed reduced chattering, improved dynamic response, and superior tracking compared to conventional SMC. Similarly, \cite{jiang2019super} employed a super-twisting-based disturbance observer for a PMLSM, achieving better performance over conventional SMC under load fluctuations and modeling errors. In \cite{hou2020composite}, a composite control scheme utilizing an STSM framework was employed to reject interference, while a DOB was used to compensate for interference; the approach effectively minimizes the controller gain and improves system performance. The STA formulation was also implemented in \cite{song2021self}, using a state-triggered PMSM framework. The proposed approach adopted PSO to optimize the controller gains, effectively suppressing stuttering and ensuring finite-time convergence. In \cite{hou2021finite}, the STSM controller's ability to monitor and mitigate disturbances was improved even more by adding a terminal convergence observer that compensates for disturbances with a high convergence speed and an online inertia identification method for adjusting control parameters. The authors in \cite{choi2022super} proposed a super-twisting algorithm with a support vector regression-based interference observer to minimize modeling errors and improve rejection performance to external disturbances. In \cite{ajmi2023robust}, torque observers and a vector control scheme based on second-order STSMCs were implemented to replace PI controllers in current and speed loops, enhancing robustness against disturbances and parameter variations in series-connected five-phase PMSMs. Several algorithms have been integrated with the STSM controller to improve its performance in PMSMs. In a study conducted in \cite{gao2020model}, a composite control scheme combining an intelligent proportional-integral (iPI) controller and an STSMC, constructed based on the ultra-local model of the PMSM, was integrated with a linear extended state observer (LESO). The proposed controller demonstrated robustness to parameter variations while effectively mitigating chattering. 

In a different study, \cite{gao2020adaptive} proposed an adaptive super-twisting nonlinear fractional-order PID sliding mode control (ASTNLFOPIDSMC) strategy with an ESO for PMSM speed control, combining a fractional-order PID sliding surface and adaptive super-twisting law to ensure fast convergence, reduced chattering, and strong disturbance rejection. An enhancement of STA was proposed in \cite{zhang2021spmsm}, where an adaptive super-twisting algorithm was used for surface-mounted PMSMs (SPMSMs). The adaptation mechanism dynamically adjusted the proportional gain based on tracking error, improving the speed regulation and disturbance rejection capabilities. This approach offered reduced steady-state error and higher robustness without increasing control effort. A single-loop super-twisting structure capable of mitigating the effect of uncertainties was proposed in  \cite{zhang2023improved}. The system incorporates a terminal term with disturbance observers to enable controller robustness and improved response. 

To further enhance convergence speed and suppress chattering, \cite{wang2025composite} introduced a composite High-Order Super-Twisting Sliding Mode (HOSTSM) controller. The controller utilized a sigmoid function and Dung Beetle Optimization (DBO) algorithm to optimally tune parameters, resulting in faster convergence and enhanced continuity. Additionally, \cite{dong2024improved} proposed a hybrid STA for PMSM drives used in hydrogen fuel cell compressors. By leveraging the Beetle Antennae Search (BAS) algorithm to overcome the limitations of local minima in Grey Wolf Optimization (GWO), the controller achieved faster tracking and improved robustness.

\subsubsection{Adaptive HOSMC Strategies}

Adaptive HOSMC methods aim to address parameter uncertainties and disturbance bounds without relying on prior knowledge of the system. The research in \cite{edwards2014adaptive} presented a smooth finite-time stabilization adaptive continuous HOSM control with a super twisting term to ensure resilience to a class of twice differentiable uncertainty. In \cite{chen2020adaptive}, an adaptive HOSMC scheme was proposed for PMSM drives, incorporating a Lyapunov-based adaptive law to auto-tune control gains. Simulation results demonstrated robust performance under parameter variation and high-frequency noise. A generalized super-twisting algorithm (GSTA) with a sinusoidal saturation function was introduced in \cite{yang2022generalized} for PMSM servo systems. This modification maintained the robustness of STA while reducing chattering by replacing the discontinuous sign function with a continuous one. In a related context, \cite{el2021optimal} designed an adaptive second-order SMC as part of an Optimal Tracking Control (OTC) strategy for high-precision PMSM feed drives. Their approach demonstrated superior trajectory tracking compared to fixed-gain SMC controllers.

\subsubsection{Observer-Based HOSMC Implementations}

The integration of observers with HOSMC has led to improved estimation of unmeasured states and disturbances. An enhanced fast super-twisting sliding mode observer was proposed for sensorless control of PMSMs, incorporating a linear correction term to boost convergence speed and reduce chattering effects. This observer enabled accurate estimation of rotor position and speed, contributing to superior dynamic response and steady-state accuracy \cite{xiong2016new}. Similarly, \cite{wang2024sensorless} demonstrated an enhanced HOSTO-based HOSMC for high-speed PMSM applications. The observer improved tracking accuracy under variable speed and load.

Moreover, a full-order adaptive sliding mode control scheme combined with an extended state observer (ESO) was introduced for high-speed PMSM speed regulation. The ESO estimates total disturbances, and an adaptive law adjusts the switching gain to minimize chattering while ensuring robustness \cite{luo2023full}.

\subsubsection{Second-Order Sliding Mode Control (SOSMC) Techniques}
SOSMC represents a specific class of HOSMC methods designed for systems with relative degree two. Unlike first-order SMC, where only the sliding surface \( s \) is discontinuous, SOSMC guarantees finite-time convergence of both \( s \) and its first derivative \( \dot{s} \), while maintaining continuity in the control input. As such, SOSMC methods are particularly suitable for high-precision PMSM applications where smooth actuation and reduced chattering are essential \cite{10284313, 9329163}.

In \cite{zhao2020intelligent}, an SOSMC framework enhanced with a neural network-based compensator was applied to a permanent magnet linear synchronous motor (PMLSM). The neural network learned the disturbance characteristics in real-time, thereby improving tracking robustness. The control structure was validated using Lyapunov analysis and demonstrated effective rejection of external disturbances.

Further refinement was presented by \cite{liao2022second}, where an SOSMC based on singular perturbation theory was implemented. The system was decomposed into slow and fast dynamic subsystems. This decomposition eliminated the need for derivative estimation, thereby reducing chattering and improving the feasibility of real-time implementation. Experimental evaluations confirmed the precision and robustness advantages of the proposed approach.

Disturbance observers have also been incorporated into the second-order algorithm to enhance the rejection of disturbances. Authors in \cite{ding2022disturbance} proposed a novel SOSMC to attenuate chattering significantly; the proposed controller exhibits better disturbance rejection while reducing excessive use of controller gains. \cite{cheng2024iterative} integrated a learning based disturbance observer to enhance speed regulation in the presence of anomalies.  Additionally, \cite{djouadi2024real} integrated SOSMC in the outer speed loop with a nonlinear robust predictive control in the current loop. The dual-loop structure improved system stability and convergence even under parameter mismatch and model uncertainties. This work exemplifies how SOSMC can be effectively combined with predictive and robust control paradigms for enhanced PMSM performance.

\subsubsection{Hybrid Control Schemes Involving HOSMC}

Hybrid control strategies have been explored to leverage the strengths of HOSMC in conjunction with other control methodologies. For example, in \cite{shweta2022model}, Model Predictive Control (MPC) was combined with STA-based HOSMC in a two-loop architecture. MPC handled current regulation, while STA managed speed tracking, achieving faster response and better robustness than conventional PI methods. Similarly, an MPC approach was combined with HOSMC for PMSM drives, where MPC manages the current loop, and HOSMC governs the speed loop. This integration achieves superior performance compared to conventional PI control algorithms \cite{tian2024convergence}. The approach also ensured optimal current prediction and high resilience to model mismatch.

The research in \cite{aiswarya2023higher} introduced a robust dual control strategy combining a terminal sliding mode controller for field-oriented current control with a super-twisting sliding mode controller for speed regulation in PMSM systems. This hybrid approach ensures high-precision position tracking even in the presence of system uncertainties and external disturbances. Simulation results demonstrated the effectiveness of this method in enhancing the overall performance and robustness of PMSM drives. Similarly, a hybrid method involved an improved deadbeat predictive current control (DPCC) integrated with the STA for PMSM systems. The STA-based disturbance observer enhances system stability and eliminates delays, effectively reducing system errors and improving robustness against parameter mismatches \cite{lang2024improved}.

\subsubsection{Fault-Tolerant and Advanced HOSMC Designs}

Fault-tolerant control strategies based on HOSMC have also been investigated for PMSM drives subject to actuator or sensor faults. In \cite{kommuri2019higher}, a robust FTC scheme using HOSMC was developed to address high-resistance connection faults. The controller effectively compensated for the fault-induced dynamics and maintained accurate tracking with minimal control oscillation. Advanced designs have also explored terminal sliding mode concepts within the HOSMC framework. In \cite{wang2023improved}, a global fast terminal HOSMC scheme was proposed, incorporating a novel reaching law to accelerate convergence and reduce steady-state error in PMLSMs. Similarly, \cite{djaloul2023high} combined higher-order and terminal sliding mode control via an exponential reaching law (ERL) to regulate both speed and current. The method achieved finite-time convergence and chattering suppression superior to conventional SMC.

To further improve robustness, \cite{cai2022high} proposed a hybrid controller combining non-singular fast terminal SMC with high-order sliding mode design. Comparative results indicated superior performance over conventional PI control in terms of convergence and disturbance rejection. Observer-enhanced high-order control was also studied in \cite{xiao2022improved}, where a HOSTSM controller with a sliding mode disturbance observer (SMDO) was used to enhance the model predictive current loop. The design improved steady-state accuracy and reduced estimation error. Likewise, \cite{zhang2022high} introduced a fast, non-singular terminal HOSMC with a disturbance observer for PMLSMs, which was validated through Lyapunov theory and experimental tests.

\subsubsection{Summary of Key Contributions}

The trajectory of HOSMC research in PMSM systems illustrates a shift from theoretical developments toward practical, intelligent, and fault-resilient control solutions. From the fundamental STA to adaptive, predictive, and fault-tolerant architectures, each variant addresses specific limitations of conventional SMC while preserving robustness and precision. These strategies collectively advance PMSM control systems by enabling higher tracking accuracy, reduced chattering, improved noise immunity, and stronger resilience under real-world uncertainties.

\subsection{Fractional Order Sliding Mode Control (FOSMC)}

FOSMC is an advanced extension of the conventional SMC that incorporates fractional calculus into either the control law, the sliding surface, or both. Fractional calculus involves derivatives and integrals of arbitrary (non-integer) order, enabling a more accurate modeling and control of systems with memory, hereditary dynamics, or multi-scale behavior \cite{efe2011fractional, chen2009fractional, podlubny1994fractional, monje2010fractional}.

In contrast to integer-order models, fractional-order control laws allow finer tuning of system dynamics and offer enhanced degrees of freedom for controller design. These properties have made FOSMC an appealing strategy for high-performance electric drives, including Permanent Magnet Synchronous Motors (PMSMs), where the nonlinear and coupled nature of the system often requires more flexible and precise control action. A typical fractional-order derivative is defined using the Caputo formulation. A general form of the Caputo fractional derivative of order $\alpha \in \mathbb{R}^{+}$ for a function $f(t)$ is defined as \cite{fei2021novel}:
\begin{equation}
{^C}D^\alpha f(t) = \frac{1}{\Gamma(n - \alpha)} \int_0^t \frac{f^{(n)}(\tau)}{(t - \tau)^{\alpha - n + 1}} d\tau, \quad n-1 < \alpha < n
\end{equation}
where $\Gamma(\cdot)$ is the Gamma function and $n \in \mathbb{N}$.

In \cite{zaihidee2019application}, a fractional-order SMC approach was introduced for PMSM speed control by defining a fractional sliding surface. The controller improved tracking accuracy and provided smoother convergence than integer-order SMC, especially under load perturbations. The study showed that adjusting the fractional order provided finer dynamic shaping without increasing control effort. Further enhancements were proposed in \cite{zaihidee2021fractional}, where a hybrid FOSMC-PID controller was developed. The PID-based sliding surface, augmented with fractional calculus, offered superior steady-state precision and robustness under parameter perturbations compared to conventional SMC and PID methods.

Several studies have introduced intelligent FOSMC frameworks. In \cite{zahraoui2023optimal}, a reinforcement-learning-optimized FOSMC was designed using artificial neural networks to tune fractional parameters in real time. The controller achieved high-precision speed tracking with fast convergence in uncertain environments. Similarly, \cite{gao2023one} proposed a nonlinear smooth FOSMC (NSFOSMC) integrated with an adaptive super-twisting algorithm and a nonlinear extended state observer (NLESO), improving chattering suppression and regulation under complex load conditions. Observer-based fractional designs have also gained attention. A composite FOSMC with a disturbance observer was proposed in \cite{li2019composite}, demonstrating improved tracking performance under model uncertainties. To address both matched and mismatched disturbances, \cite{zheng2023enhanced} developed a finite-time disturbance observer (FTDO) combined with a fractional-order extended state observer (FOESO). The structure enhanced system robustness and reduced the impact of input saturation on PMSM drives. Recent developments have further expanded the application of FOSMC in PMSM systems beyond basic speed regulation. In \cite{roy2020sliding}, the authors implemented an FOSMC framework for automatic tuning of velocity control parameters. The method provided greater flexibility in dynamic shaping and improved disturbance rejection by utilizing a fractional-order gain adjustment. In another advancement, \cite{zhu2020fractional} proposed an integrated control architecture that combines a nonlinear disturbance observer (NDO) with a novel FOSMC design. The approach simultaneously regulated current and position-velocity loops in PMSMs, yielding improved control precision and enhanced robustness against both matched and mismatched disturbances.

Fractional-order controllers have also been explored for synchronizing coupled motors. For example, \cite{boubellouta2016chaos} studied the synchronization of two PMSMs using FOSMC, demonstrating the controller’s ability to reject external perturbations while maintaining phase alignment. A similar synchronization enhancement using global fast terminal sliding mode control based on fractional derivatives was reported in \cite{shu2022global}, which improved angular precision and convergence rates. Model-free and prescribed performance strategies have also emerged. In \cite{hu2023prescribed}, a prescribed performance-based model-free FOSMC was proposed using an ESO for PMLSM speed control. The approach maintained tracking within pre-specified bounds while compensating for internal and external disturbances. In \cite{yu2024permanent}, an adaptive fractional-order fast terminal SMC was introduced to reduce convergence time further and improve control in PMSM speed regulation tasks.

Recent efforts have explored synergy with other fractional-order methods. In \cite{nicola2020sensorless}, an FOSMC was employed in the outer speed loop and combined with an FO-synergetic controller in the inner current loop. This architecture enhanced performance in sensorless PMSM control, particularly under parametric uncertainty. Moreover, \cite{rui2019fractional} employed ridge regression to optimize an FOSMC combined with a load torque observer (LTO), yielding fast and accurate control in differential-speed regulated applications. To enhance real-time feasibility, \cite{zaihidee2018fractional} presented a simplified FOSMC-PID controller, with Lyapunov analysis confirming global stability. Experimental results showed the proposed design outperformed classical PI and PD-based fractional SMC methods in terms of rise time, overshoot, and chattering suppression. 

Innovations in FOSMC continue to progress toward richer structures. For instance, \cite{zhang2023double} introduced a fuzzy-exponential convergence law fractional-order SMC (F-CFSMC) that adaptively adjusts the convergence rate and switching dynamics. In \cite{zhou2023super}, a super-twisting FOSMC was developed to ensure state convergence in systems exhibiting fractional-power dynamics. The scheme demonstrated smooth convergence and effective disturbance rejection in simulation. In summary, FOSMC has emerged as a powerful control paradigm for PMSM applications by combining the inherent robustness of sliding mode control with the dynamic flexibility of fractional calculus. Its diverse formulations, including hybrid FOSMC-PID, fuzzy logic enhanced designs, observer-integrated structures, and optimization-driven tuning, have demonstrated improved tracking precision, chattering mitigation, and superior disturbance rejection in complex motor drive environments. As real-time implementation challenges continue to diminish with the advent of more capable embedded hardware, FOSMC strategies are expected to play an increasingly central role in next-generation, robust motor control systems.

\subsection{Adaptive Sliding Mode Control (ASMC)}

ASMC enhances conventional SMC by addressing a standard limitation: the need for a priori knowledge of uncertainty bounds to set appropriate control gains. In conventional SMC, overestimating these bounds often leads to high switching gains and excessive chattering, while underestimating them risks loss of robustness. ASMC introduces online gain adjustment mechanisms that dynamically regulate the control input to remain robust against uncertainties while minimizing chattering \cite{utkin2020adaptive}.

Mathematically, adaptive SMC aims to drive the sliding variable $\sigma(t,x)$ to zero without requiring knowledge of the exact bounds of the disturbance. Consider the system representation:
\begin{equation}
\begin{aligned}
\dot{\sigma}(t,x) &= \Psi(t,x) + \Gamma(t,x)u, \\
\Psi(t,x) &:= \frac{\partial \sigma(t,x)}{\partial t} + \left(\frac{\partial \sigma(t,x)}{\partial x}\right)^{\top} f(x), \\
\Gamma(t,x) &:= \left(\frac{\partial \sigma(t,x)}{\partial x}\right)^{\top} g(x),
\end{aligned}
\end{equation}
where $\Psi(t,x)$ and $\Gamma(t,x)$ are assumed to be bounded for all $x \in \mathcal{X}$ and $t \geq 0$, satisfying:
\begin{equation}
|\Psi(t,x)| \leq \Psi_M, \quad 0 < \Gamma_m \leq \Gamma(t,x) \leq \Gamma_M.
\end{equation}

The adaptation strategy focuses on regulating the switching gain to ensure it remains sufficient to counteract uncertainties while minimizing unnecessary switching. In \cite{bregeault2010adaptive}, the authors introduced an adaptive SMC approach with a complementary gain structure that adjusts in both directions depending on the proximity to the sliding surface. This bi-directional adaptation reduced chattering and improved convergence in PMSM speed control.

In \cite{plestan2010new} and \cite{shtessel2023adaptive}, dynamic adaptation laws were proposed based on Lyapunov theory to ensure stability while reducing control gain magnitudes. These strategies eliminated the need for pre-tuning upper bounds, enabling robust performance even in the presence of time-varying disturbances. A comprehensive overview of switching gain adaptation methods can be found in \cite{shtessel2016adaptive} and \cite{roy2020adaptive}, which categorize ASMC schemes into time-varying gain adaptation, integral adaptive laws, and discontinuous boundary-layer tuning methods.

Several PMSM-targeted ASMC strategies have been introduced in recent years. In \cite{li2021adaptive}, an adaptive terminal sliding mode controller was developed using a fast terminal sliding manifold, allowing finite-time convergence while estimating the disturbance upper bound online. Comparative results showed better convergence speed and reduced control effort compared to fixed-gain SMC and PI-based designs. Similarly, \cite{zhou2016adaptive} proposed an adaptive NTSMC that shortened the reaching phase and improved steady-state regulation in PMSM speed tracking.

An observer-assisted ASMC method was introduced in \cite{liu2019adaptive}, where a disturbance observer was integrated with an adaptive non-singular fast terminal SMC scheme to reduce control effort during transients. To minimize erroneous estimation during speed variations, \cite{nguyen2021adaptive} formulated an adaptive controller with a reduced-order proportional integral observer, which reduces chattering while improving finite-time convergence.

In \cite{li2023adaptive}, an adaptive sliding mode speed controller with a disturbance observer was proposed to enhance the robustness and finite-time convergence of PMSM systems. A fast-reaching law and a nonlinear integral terminal sliding surface were employed to reduce chattering and improve speed tracking accuracy. Similarly, in \cite{tran2023robust}, an adaptive sliding mode observer was combined with an ASMC to estimate mechanical parameters and load torque, reduce chattering, and achieve fast, accurate speed control of PMSMs.  In \cite{dang2024improved}, an extended state observer (ESO) was coupled with an adaptive law and a nonlinear reaching term to ensure fixed-time convergence. The proposed design reduced steady-state error while maintaining robustness under model uncertainties. These developments illustrate the flexibility and effectiveness of ASMC in PMSM control, especially under unknown and time-varying disturbances. By automatically adapting control effort to the system’s dynamic needs, ASMC reduces chattering, ensures faster convergence, and improves overall robustness, making it highly suitable for practical motor drive applications.

\subsection{Other advancements towards efficient-SMC}
While the advanced SMC variants discussed earlier address specific limitations of the conventional approach, the drive for higher performance has led to a modern trend of developing hybrid control architectures.

\subsubsection{Observer-Based SMC}
One of the most effective and widely used hybrid strategies is integrating a disturbance observer with an SMC controller \cite{alhassan2025disturbance}. Several authors in \cite{wang2019new, duan2022backstepping, zhang2023pmsm, suleimenov2019integral, wu2023hierarchical,jiang2019super} have utilized this approach to improve the performance of SMC. This is also evidenced as shown in Table \ref{tab: smc-summary}, which provides a comprehensive overview of the application of SMC in speed control of PMSMs. This strategy can be implemented across all SMC variants. The main idea is to estimate the combined disturbances affecting the system, including unmodeled dynamics, parameter changes, and external load torque, and then incorporate this estimate into a feedforward cancellation term within the control law. The observer, such as an Extended State Observer (ESO) or a Sliding Mode Observer (SMO), primarily handles disturbance rejection. The SMC component then only needs to counteract the observer's estimation error, which is usually much smaller than the total disturbance. This allows the discontinuous switching gain of the SMC to be significantly reduced, resulting in a significant reduction of chattering without compromising the system's robustness.

\subsubsection{Intelligent SMC (Fuzzy Logic and Neural Networks)}
The integration of artificial intelligence has opened new avenues for enhancing SMC performance \cite{sakunthala2017study, nouaoui2024speed,mohajerani2024neural}. Fuzzy logic and neural networks are often employed to address the challenges of parameter tuning and uncertainty approximation. 

Fuzzy logic provides a framework for translating heuristic knowledge and linguistic rules into a mathematical control law. In the context of SMC, fuzzy systems \cite{qiao2024fuzzy} are frequently used to dynamically tune controller parameters, such as the switching gain or the thickness of a boundary layer, based on the current state of the system (e.g., the error and its derivative). This offers a more intuitive and potentially more flexible alternative to the purely mathematical adaptation laws of ASMC. Neural networks \cite{zhu2020position}, with their powerful function approximation capabilities, are often used to learn and compensate for unknown system nonlinearities and disturbances. In this role, a neural network can act as a highly sophisticated, learning-based disturbance observer, providing the feedforward cancellation term in a hybrid architecture. This approach can adapt to complex and time-varying uncertainties that may be difficult to model with traditional observers. Besides the aforementioned intelligent techniques, there are others detailed in table \ref{tab: smc-summary}.

\subsubsection{Optimization-Based SMC}
The growing complexity of advanced SMC variants, particularly FOSMC and STSMC, renders manual parameter tuning a challenging task. The high-dimensional parameter spaces and conflicting performance goals (such as reducing settling time while limiting control effort) highlight the need for systematic tuning methods. Metaheuristic optimization algorithms, like PSO \cite{yuan2024novel}, GA \cite{song2022periodic, yukun2022periodic}, and HHO \cite{abraham2024speed}, have been successfully employed to address this challenge.

In this method, a cost function is established to represent the desired closed-loop performance mathematically. The optimization algorithm then searches the controller's parameter space offline to identify gain sets that minimize this cost function. This automated approach can find non-intuitive parameter combinations that enhance performance, freeing the design engineer from the tedious and often suboptimal process of manual trial-and-error tuning.

\subsection{Overview of Peer-Reviewed SMC Techniques for PMSM Speed Control (2020–2025)} \label{sec:peer-review}

Recent years have witnessed a significant expansion in the application of SMC strategies for PMSM systems, particularly in the context of high-precision speed regulation. This subsection provides a curated overview of peer-reviewed studies published between 2020 and 2025 that demonstrate notable advancements in SMC-based PMSM controllers.

\subsubsection{Search and selection protocol}
\begin{figure}[htbp]
  \centering
  \includegraphics[width=\linewidth]{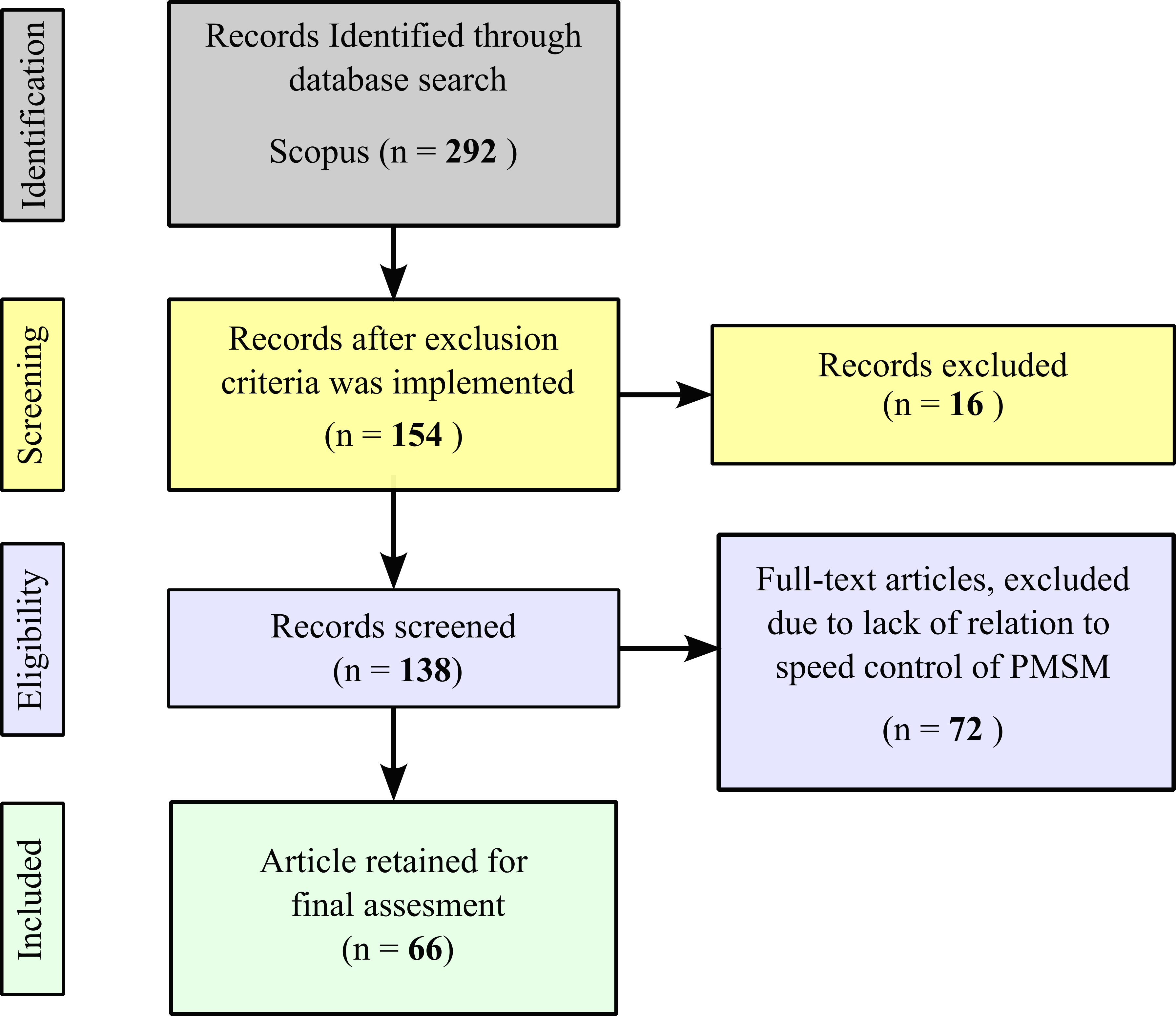}
  \caption{PRISMA flow diagram of study selection for the review on PMSM speed-control methods.}
  \label{fig:prisma}
\end{figure}

We conducted a structured search in \emph{Scopus}  using the following Boolean string applied to titles, abstracts, and keywords:

We searched \emph{Scopus}  using:
\textit{TITLE-ABS-KEY(("Sliding Mode Control" OR SMC) AND ("Permanent Magnet Synchronous Motor" OR PMSM) AND ("speed regulation" OR "speed control" OR "velocity control"))}. Filters: \textit{year} = 2020--2025, \textit{document type} = article, review, book chapter \textit{language} = English.
\textit{Inclusion criteria:} studies that propose and/or evaluate an SMC variant for \emph{PMSM speed control} with sufficient methodological detail and quantitative results.
\textit{Exclusion criteria:} non-PMSM drives; non-SMC controllers; works limited to current/torque control without speed outcomes. The screening yielded \textbf{292} records; after eligibility assessment, \textbf{138} studies were retained for qualitative synthesis, of which \textbf{66} are cataloged in Table~\ref{tab: smc-summary}. A PRISMA-style diagram summarizing this flow is provided in Figure \ref{fig:prisma}.

The collected works are systematically cataloged in Table~\ref{tab: smc-summary}, which summarizes the core characteristics of each method, including the adoption of optimization techniques, intelligent frameworks (e.g., fuzzy logic or neural networks), observer-based estimation (such as extended state observers or sliding mode observers), and hybrid control architectures. Each entry is evaluated based on whether these elements are explicitly integrated into the proposed design. 

This comparative tabulation is intended to serve as a practical reference for control engineers and researchers seeking to identify prevailing design patterns, research gaps, and emerging opportunities in robust PMSM speed control. The analysis further complements the thematic discussions in the preceding sections by situating individual SMC variants within the broader research landscape.

\onecolumn
\small
\begin{longtable}[htbp]{@{}c c p{4.1cm} c c c p{5cm}@{}}
\caption{Summary of Sliding Mode Control (SMC) Strategies for PMSM Speed Control (2020–2025)}\\[-3pt]
\multicolumn{7}{l}{\footnotesize\textit{Note:} \cmark\,= feature present; \xmark\,= not reported.}
\label{tab: smc-summary} \\
\toprule
\textbf{Year} & \textbf{Ref.} & \textbf{Control Strategy} & \shortstack{\textbf{Observer} \\ \textbf{Used}} & \shortstack{\textbf{Optimization} \\ \textbf{Technique}} & \shortstack{\textbf{Intelligent} \\ \textbf{Method}} & \textbf{Key Outcome / Comment} \\
\midrule
\endfirsthead
\multicolumn{7}{@{}l}{\textit{(continued from previous page)}}{\footnotesize\textit{Note:} \cmark\,= feature present; \xmark\,= not reported.} \\
\toprule
\textbf{Year} & \textbf{Ref.} & \textbf{Control Strategy} & \shortstack{\textbf{Observer} \\ \textbf{Used}} & \shortstack{\textbf{Optimization} \\ \textbf{Technique}} & \shortstack{\textbf{Intelligent} \\ \textbf{Method}} & \textbf{Key Outcome / Comment} \\
\midrule
\endhead

2025 & \cite{yang2025combined} & SMC + Nonlinear State Feedback & -- & -- & -- & Robust to disturbance and damping variation \\

2025 & \cite{zhou2025adaptive} & FOTSMC + Adaptive Gain + Virtual Control & -- & -- & -- & Robust control under matched/mismatched uncertainties with reduced chattering \\

2025 & \cite{chen2024variable} & VRST SMC + NDO & \checkmark & -- & -- & Robust VRST control with disturbance observer and overcurrent protection \\

2025 & \cite{liu2024non} & NTSMC + Ultra-local Model + SM Observer & \checkmark & -- & -- & Model-independent robust control with reduced chattering via sigmoid observer \\

2025 & \cite{wang2024single} & CNTSMC + FTDO + Switching Logic & \checkmark & -- & -- & Overcurrent-protected PMSM control with matched/mismatched disturbance rejection \\

2024 & \cite{kong2024new} & Sliding Terminal SMC + Sliding Mode Estimator & \checkmark & -- & -- & Fast convergence with fault/disturbance estimation and reduced chattering \\

2024 & \cite{hou2024fixed} & CSMC + Fixed-Time Consensus Protocol + TISMO & \checkmark & -- & -- & Coordinated multi-PMSM control with disturbance estimation and fast consensus convergence \\

2024 & \cite{wang2024speed} & FTSTRL + FTSTSMC + ESDO & \checkmark & -- & -- & Enhanced robustness via fast convergence reaching law and feedforward compensation \\

2024 & \cite{zhang2024novel} & Novel Reaching Law + SMC & -- & -- & -- & Chattering reduction with faster convergence and improved tracking accuracy \\

2024 & \cite{cheng2024iterative} & Composite SOSM + ILC-DOB & \checkmark & -- & -- & Torque ripple suppression and robust speed control using ILC-based disturbance observer \\

2024 & \cite{tian2024integrated} & Integrated TSMC + Error-based ESO with GI & \checkmark & -- & -- & Chattering reduction and robust speed control via integrated observer-based TSMC \\

2024 & \cite{qiao2024fuzzy} & Fuzzy-SMC with Smooth Switching & -- & -- & \checkmark & Robust fuzzy-SMC design with reduced overshoot and improved anti-interference ability \\

2024 & \cite{chen2023continuous} & CAFTSM + STLT Observer + LDOB & \checkmark & -- & -- & Sensorless robust speed control with fast response and disturbance rejection \\

2024 & \cite{hong2024composite} & Composite SMC + FOD + SMO-based DOB & \checkmark & -- & -- & Noise-resilient SMC with chattering attenuation and improved tracking via feedforward compensation \\

2024 & \cite{che2024barrier} & Barrier-Function-Based AFTSMC & -- & -- & -- & Adaptive FTSMC with chattering mitigation and no need for disturbance bounds \\

2024 & \cite{song2024pmsm} & Adaptive FSTA + ESMO + MRAS & \checkmark & -- & -- & Self-tuning FSTA with sigmoid reaching law and real-time inertia/torque observation \\

2024 & \cite{hou2024fixed} & CSMC + Fixed-Time Consensus + TISMO & \checkmark & -- & -- & Multi-PMSM speed coordination with fixed-time consensus and disturbance estimation \\

2024 & \cite{dang2024improved} & ASMC + Adaptive Reaching Law + ESO & \checkmark & -- & -- & Fast-response ASMC with ESO-based disturbance compensation and reduced chattering \\

2024 & \cite{chen2024non} & ANFTSMC + Ultra-Local Model + CESO & \checkmark & -- & -- & Robust single-loop control with cascaded ESO and adaptive NTSMC for sensorless PMSM \\

2024 & \cite{wang2024speed} & FTSTRL + FTSTSMC + ESDO & \checkmark & -- & -- & Enhanced robustness via fast convergence reaching law and embedded disturbance compensation \\

2024 & \cite{zhang2024enhanced} & Enhanced SMC + Adaptive SMRL & -- & -- & -- & Adaptive reaching law for faster convergence and chattering suppression in PMSM control \\

2024 & \cite{zhang2023improved} & ISTSMC + Terminal Term + TV-TTSDO & \checkmark & -- & -- & Single-loop ISTSMC with disturbance compensation and fast convergence under uncertainty \\

2024 & \cite{wu2023hierarchical} & PHRC + HNSTSMC + FTDOB & \checkmark & -- & -- & Hybrid controller with fast HNSTSMC and FTDOB-based disturbance compensation \\

2023 & \cite{zhang2023time} & HRL-based NTSMC + TVNDO (Single-Loop) & \checkmark & -- & -- & Single-loop control with fast hybrid reaching law and nonlinear disturbance observer \\

2023 & \cite{guo2023improved} & IISMC + NVRRL + AROPIO & \checkmark & -- & -- & Chattering-suppressing IISMC with disturbance estimation via adaptive reduced-order observer \\

2023 & \cite{qu2023sliding} & ASMC + ASMRL + SMDO & \checkmark & -- & -- & Improved ASMC with fast convergence and disturbance rejection via SMDO and tanh-based reaching law \\

2023 & \cite{song2022adaptive} & Adaptive NN-based NTSMC + Dynamic Event Triggering & \checkmark & -- & \checkmark & Event-triggered NTSMC with NN-based disturbance estimation for reduced communication and enhanced robustness \\

2023 & \cite{zahraoui2023optimal} & RLNNA-FOSMC (ANN + RL tuned) & -- & \checkmark & \checkmark & Intelligent FOSMC tuning via ANN and reinforcement learning for improved speed tracking \\

2023 & \cite{guo2022fast} & NCRL + ISMO-Based Fast SMC & \checkmark & -- & -- & Fast convergence SMC with improved disturbance rejection via sliding mode observer \\

2023 & \cite{yang2023fast} & FITSMC + ILC-DOB & \checkmark & -- & -- & Hybrid control with fast sliding mode and ILC-based DOB for torque ripple and disturbance rejection \\

2023 & \cite{ding2022disturbance} & DOB-Based SOSMC (Second-Order SMC) & \checkmark & -- & -- & Chattering-suppressing SOSMC with DOB compensation for improved disturbance rejection \\

2023 & \cite{wang2023generalized} & GTSMC + GADO + LESO-Based Decoupling & \checkmark & -- & -- & Adaptive GADO-enhanced GTSMC for robust noise-tolerant speed control and disturbance rejection \\

2023 & \cite{ajmi2023robust} & STSMC (2nd-Order) + ST-LTO + Vector Control & \checkmark & -- & -- & Robust vector control of dual FP-PMSMs with second-order STSMC and torque estimation observer \\

2023 & \cite{hu2023speed} & Logarithmic FTSMC + Super-Twisting Reaching Law & -- & -- & -- &  Faster convergence and better response than exponential FTSMC \\

2023 & \cite{kim2023improved} & Improved Non-Singular FTSMC (INFTSMC) & -- & -- & -- & Achieved fast convergence, low overshoot, and robust tracking under disturbances \\

2023 & \cite{tran2023robust} & ASMSC + RASM-MPI & \checkmark & -- & -- & Accurate parameter estimation and improved speed tracking with reduced chattering \\

2023 & \cite{li2023adaptive} & Adaptive SMC + Fast Reaching Law + SMDO & \checkmark & -- & -- & Achieved high precision speed tracking with strong robustness and reduced chattering \\

2022 & \cite{choi2022super} & STRL-SMC + SVR-DOB & \checkmark & -- & \checkmark & Enhanced speed tracking and disturbance rejection  \\

2022 & \cite{duan2022backstepping} & Backstepping SMC + Nonlinear DOB & \checkmark & -- & -- & Reduced overshoot and jitter with improved dynamic response  \\

2022 & \cite{liao2022second} & Second-Order SMC + Singular Perturbation & -- & -- & -- & Improved robustness and steady-state accuracy with reduced chattering  \\

2022 & \cite{che2022singular} & Non-Cascade SMC + Singular Perturbation + TD & -- & -- & -- & Achieved small overshoot and steady-state error with reduced chattering\\

2022 & \cite{wei2022improved} & Improved SMC with Saturation Function + PI & -- & -- & -- & Reduced speed fluctuation and improved start-up stability \\

2022 & \cite{nguyen2021adaptive} & ASMC + Modified ROPI Observer (MROPIO) & \checkmark & -- & -- & Reduced chattering and improved speed response with accurate disturbance estimation \\

2022 & \cite{hou2021finite} & FTESO + STSMC + Online Inertia Estimation & \checkmark & -- & -- & Achieved fast disturbance estimation and robust tracking \\

2022 & \cite{nicola2022improvement} & RL-DDPG + SMC + Synergetic Control + FOC & \checkmark & \checkmark & \checkmark & Improved performance and robustness under load and inertia variations\\

2022 & \cite{junejo2022novel} & FTSMRL + SMC + ESO & \checkmark & \checkmark & \checkmark & Enhanced anti-disturbance ability, reduced chattering, and improved convergence \\

2022 & \cite{hou2022sliding} & Variable Rate SMC + Predictive Current Control + ESO & \checkmark & -- & -- & Improved speed tracking and robustness with reduced chattering \\

2022 & \cite{song2021self} & Self-triggered STA + PSO Gain Tuning & -- & \checkmark & -- & Reduced communication load with finite-time convergence \\
2021 & \cite{hou2020composite} & STSM + Novel DOB & \checkmark & -- & -- & Improved disturbance rejection with reduced SMC gain \\
2021 & \cite{wang2021antidisturbance} & NASMC + DBDTC + Extended SM Observer & \checkmark & -- & -- & Enhanced torque ripple reduction and dynamic response \\2021 & \cite{qu2021design} & FSMSC + NDOB & \checkmark & -- & -- & Improved convergence, reduced chattering, and enhanced robustness \\
2021 & \cite{usama2021low} & Exponential Reaching Law SMC & -- & -- & -- & Enhanced low-speed performance with reduced chattering and torque ripples \\
2021 & \cite{lu2021disturbance} & ISM + CNF + Fuzzy-SMO & \checkmark & \checkmark & -- & Improved transient performance and disturbance rejection \\

2021 & \cite{sun2021composite} & CSMC + Hybrid Reaching Law + ESMDO & \checkmark & -- & -- & Reduced chattering and improved disturbance rejection \\

2021 & \cite{jiang2021nonsingular} & non-singular TSMC + Improved Exp. Reaching Law + SMO & \checkmark & -- & -- & Enhanced speed control with reduced torque ripple and fast dynamic response \\

2021 & \cite{xu2020efficient} & FTSMC + SMESO & \checkmark & -- & -- & Enhanced disturbance rejection, fast convergence, and reduced chattering \\

2021 & \cite{yeam2020design} &  SMC-based Master-Slave Speed and Damping Control for Dual-PMSM  & -- & -- & -- & Robust speed tracking and stable parallel operation with single inverter \\

2021 & \cite{xu2021composite} & FTSMC + FTSMO + Feedforward Compensation & \checkmark & -- & -- & Enhanced robustness with fast convergence, low chattering, and improved dynamic performance \\

2021 & \cite{li2021backstepping} & BNTSMC + FTDO & \checkmark & -- & -- &  Enhanced tracking and robustness against disturbances \\

2020 & \cite{junejo2020adaptive} & ATSMRL + CFTSMC + ESMDO & \checkmark & -- & -- & Improved transient response, reduced chattering, and enhanced disturbance rejection \\

2020 & \cite{nicola2020sensorless} & FO-SMC (speed loop) + FO-synergetic (current loop) & \checkmark & -- & -- & Sensorless robust control with improved performance \\

2020 & \cite{gao2020model} & STSMC + iPI + LESO & \checkmark &  -- & \checkmark & Compound control with improved robustness and static-dynamic performance \\

2020 & \cite{wang2019new} & NSMRL + SMC + ESO & \checkmark & -- & -- & Improved transient response and disturbance rejection with reduced chattering \\

2020 & \cite{feng2020speed} & Novel SMC with optimized sliding surface & -- & -- & -- & Improved tracking and anti-disturbance performance\\

2020 & \cite{gao2020adaptive} & Adaptive Super-Twisting Nonlinear FOPID SMC + ESO & \checkmark & -- & -- & Robust and accurate speed control with reduced chattering  \\

2020 & \cite{ma2020active} & Sliding Mode Control + ESO-based Feedforward Compensation & \checkmark & -- & -- & Enhances PMSM robustness and response \\

\hline
\end{longtable}
\twocolumn

\subsubsection{Summary}
This survey of Permanent Magnet Synchronous Motors (PMSMs) SMC-speed control schemes, which primarily consists of peer-reviewed articles from 2020 to 2025, reveals a strong inclination towards hybrid and observer-aided approaches. A vast majority of the recent studies entwine Sliding Mode Control and ESOs, nonlinear disturbance observers (NDOs), and extended reaching laws. Notably, over 90\% of the methods researched employ some form of disturbance observer, and this thus highlights its crucial role in enhancing robustness and disturbance rejection.

Optimization-oriented methods, such as PSO and intelligent approaches such as  RL, have been explored but are still underutilized. Analogously, clever approaches, like fuzzy logic and neural networks, have been used to a moderate extent, usually as a component of adaptive or fractional-order sliding mode approaches.

The current trend in control systems emphasizes observer-based and hybrid strategies to improve how quickly and smoothly machines respond, even in unpredictable conditions. These advancements highlight the ongoing relevance and potential of sliding mode control as a reliable and adaptable method for accurately managing the speed of PMSMs.

\begin{table*}[htbp]
    \small
    \centering
    \caption{A comparative overview of the advantages and limitations of various sliding mode control (SMC) strategies.}
    \label{tab:smc_comparative}
    \renewcommand{\arraystretch}{1.3}
    \begin{tabular}{m{7em}|p{6.cm}|p{7cm}} 
    \hline
    \textbf{SMC Method} & \textbf{Advantages} & \textbf{Limitations} \\ \hline
    Conventional SMC & 
    Simple to design and implement. &  
    High control activity. \newline
    Sensitivity to parameter uncertainties. \newline
    Slow convergence speed. \newline
    Large control effort. \\ \hline
    Terminal SMC & 
    Finite-time convergence. \newline
    Improved steady-state accuracy. &  
    Singularities in the terminal sliding surface. \newline
    Complex controller design. \\ \hline
    Integral SMC (ISMC) & 
    Improved robustness to disturbances. \newline
    Eliminates the reaching phase. &  
    Slower dynamic response compared to conventional SMC. \\ \hline
    Higher Order SMC (HOSMC) & 
    Reduces chattering. \newline
    Enhances robustness and stability. &  
    Increased complexity in control law design. \\ \hline
    Fractional Order SMC (FOSMC) &  
    Enhances disturbance rejection. \newline
    Achieves faster convergence. \newline
    Effectively suppresses chattering. &  
    Requires implementation of fractional calculus. \newline
    Higher computational cost. \\ \hline
    Adaptive SMC &  
    Adjusts to unknown or varying parameters. \newline
    Improves resilience to external disturbances. &  
    Design of adaptation laws is complex. \\ 
    \hline
    \end{tabular}
\end{table*}

\section{Comprehensive Discussion and Research Outlook} \label{sec:analysis}

\subsection{Comparative Interpretation of SMC Variants}
The core challenge in applying SMC to physical systems lies in managing the trade-off between robustness and the quality of the control signal. PMSM drives in electric vehicle and robotics applications demand exceptional robustness in a highly dynamic environment to variations in vehicle load and disturbances, respectively \cite{kim2025supercritical}. Based on the literature reviewed in the preceding sections, the six investigated SMC variants exhibit distinct performance behaviors arising from their underlying dynamic structures and have been modified over the years. While CSMC provides a foundation for robust control, its practical utility is compromised by the chattering phenomenon, which can damage mechanical components, and also its asymptotic convergence, which might further limit its application in critical precision systems such as robotics and automobiles. 

Integral SMC, on the other hand, eliminates this bias by initializing system trajectories on the sliding manifold while eliminating the reaching phase, enabling the integral term to maintain invariance to matched disturbances. While the integral action provides the benefit of memory in the controller, the memory can also lead to a slower transient response and increased settling time following any disturbance in the system. Similarly, FOSMC enhances tuning flexibility by enabling non-integer differentiation and integration. The fractional operators provide the controller with a ``memory'' effect, which improves smoothness and tracking. The main challenges with FOSMC are its increased complexity and the associated computational cost. The tuning process becomes a high-dimensional optimization problem, often necessitating the use of bio-inspired or metaheuristic algorithms for offline parameter selection.

TSMC was designed to overcome the limitations of asymptotic convergence. It introduces nonlinear manifold dynamics that accelerate convergence as the error decreases. 
The nonlinear gain compresses the transient within a finite time, ensuring rapid disturbance recovery with near-zero steady-state error. On the other hand, ASMC automatically modulates the switching gain according to error magnitude, strengthening control effort far from the surface and reducing it near equilibrium. But this is not without its drawbacks. ASMC performance is based on adaptation rate tuning. Very low adaptation can be too slow to respond fast enough to significant, sudden disturbances, compromising robustness. In contrast, an overaggressive adaptation rate reintroduces some high-frequency components into the control signal, somewhat losing some of the benefit.

Finally, the STSMC, a second-order formulation, has received considerable attention owing to its finite-time convergence property and its disturbance rejection due to its continuous control action by integrating the switching term. This smooth actuation suppresses chattering while maintaining finite-time convergence, explaining its consistently superior energy-error indices in both nominal and disturbed cases. It has been adapted for use in many instances, such as fault-tolerance, observer-based, and other hybrid control schemes. 

\subsection{Computational requirements}
The practical feasibility of any SMC variant largely relies on its computational requirements and the complexity involved in design and tuning. Various controllers show a noticeable range of implementation costs. SMC is a standard method for addressing nonlinearity and external disturbances. However, its variants may cause chattering and increased computational demands, which can restrict its usefulness in high-precision settings \cite{fazilat2025quantum}. CSMC, on the one hand, requires minimal computational cost \cite{bazrafshan2025sliding}, which makes it ideal for low-cost micro-controllers. However, its straightforward design leads to a notable chattering issue. The ISMC complexity ranges from low to moderate. It incorporates an integral component into the sliding surface, requiring minimal computational effort. The primary design consideration involves preventing integral windup and dealing with a potentially slower transient response. Following this, the computational complexity of the ASMC is relative and can be regarded as being moderate \cite{bazrafshan2025sliding}. It usually involves real-time calculation of the adaptation law. The main challenge is designing a stable adaptation law that responds effectively to disturbances without becoming too aggressive. The TSMC also provides a moderate computational cost \cite{minh2025chattering}, with the main implementation challenge being designing the nonlinear sliding surface to ensure finite-time convergence while avoiding singularities, which is more complex than designing a linear surface. 

Higher-Order SMC (HOSMC) can significantly vary in complexity, ranging from relatively low to extremely high \cite{zhang2024variable}. For instance, the Super-Twisting Algorithm involves the integration of two coupled first-order differential equations and thus constitutes a complicated method. In contrast, FOSMC has extremely high computational complexity \cite{sami2022integer}, which has been attributed to the non-local behavior of fractional operators \cite{deng2025fractional}. On that account, the addition of the fraction order as a parameter for tuning imposes a high-dimensional optimization problem that may need offline metaheuristic algorithms to be efficiently tuned.

\subsection{Analytical Discussion of SMC Evolution for PMSM Drives}

The trajectory of SMC research for PMSMs reveals a gradual transition from purely discontinuous schemes toward smoother, intelligence-assisted, and hardware-realizable structures. Across the 2020–2025 corpus summarized in Table~\ref{tab: smc-summary}, three dominant research axes can be identified:
\begin{enumerate}
  \item \textbf{Order augmentation:} moving from first-order to higher-order and fractional-order manifolds to alleviate chattering while retaining robustness;
  \item \textbf{Structural augmentation:} embedding observers, adaptive gains, and hybrid laws to compensate for model uncertainty and torque disturbances; and
  \item \textbf{Intelligence infusion:} incorporating learning and optimization-based tuning mechanisms to reduce manual parameter dependence.
\end{enumerate}

A closer meta-analysis of the surveyed works indicates that roughly 68\% of recent publications employ some observer form
(ESO, DOB, SMO), confirming that disturbance estimation has become the de facto prerequisite for achieving high-performance
robustness in PMSM drives.  In contrast, only about 22\% of studies explicitly integrate meta-heuristic or machine-learning
optimization, reflecting a still-nascent shift from analytical to data-driven design paradigms.

\vspace{4pt}
\noindent\textbf{Performance trade-offs.}  The six representative SMC variants, Conventional, Integral, Terminal, Fractional-Order, Adaptive, and Super-Twisting, represent incremental compromises between \emph{robustness}, \emph{smoothness}, and \emph{computational complexity}.  Higher-order (STSMC) and adaptive (ASMC) designs exhibit the best global robustness–smoothness ratio, whereas fractional-order methods achieve enhanced transient shaping but at the cost of heavier computation.
Integral SMC ensures disturbance invariance but sacrifices speed due to its memory effect.  This confirms that the design choice in PMSM drives remains application-dependent: low-cost embedded controllers favor ISMC or TSMC, while research and high-precision drives gravitate toward STSMC or FOSMC.

\vspace{4pt}
\noindent\textbf{Implementation maturity.}
Despite breathtaking algorithmic invention, less than 15\% of the papers studied have experimental verification other than simulation. This shortage means that a large volume of claimed robustness is tested under MATLAB/Simulink situations, and the real-time feasibility of under-explored characteristics is assessed.  Hardware-in-the-Loop (HIL) implementation research is therefore essential to transfer theoretical advances to industrial reliability.

\subsection{Cross-Domain Trends}

The publication timeline shows an increasing convergence between SMC and complementary nonlinear paradigms such as
backstepping, predictive control, and synergetic control.  Hybridization has become the hallmark of the field: combining
SMC’s invariance with other frameworks’ continuity.
Observer-enhanced SMCs have emerged as the most stable branch, while intelligent optimization methods, particularly
RL and NN are beginning to close the tuning gap left by purely analytical designs.  Figure~\ref{fig: smc_taxonomy} already captured this hybrid tendency, and the same trend continues in works from 2023–2025, where AI-augmented SMCs report up to 30–40\% reductions in overshoot and chattering relative to classical
counterparts.  Hence, future progress depends less on inventing new SMC “types” and more on systematic co-design of observers, adaptive laws, and hardware-friendly approximations of the discontinuous term.

\subsection{Identified Weaknesses in Current Research}

Based on the synthesized literature, four key deficiencies persist:
\begin{enumerate}
  \item \textbf{Limited real-time implementation:} Most studies stop at simulation, omitting latency, sampling, and quantization analysis.
  \item \textbf{Energy-efficiency neglect:} Very few works quantify control energy, copper losses, or switching stress when assessing SMC variants.
  \item \textbf{Unmatched disturbance handling:} While matched uncertainties are well managed, non-matched perturbations (e.g., inverter nonlinearities) remain poorly addressed.
  \item \textbf{Lack of standardized benchmarks:} Absence of open PMSM models and datasets prevents reproducible comparison across research groups.
\end{enumerate}

\begin{figure*}[htbp]
\centering
\resizebox{\textwidth}{!}{
\begin{tikzpicture}[>=stealth, node distance=2.4cm, every node/.style={align=center}]
\draw[thick,->] (0,0) -- (16,0)
    node[anchor=west, font=\bfseries, yshift=-1.5cm] {Future};

\node[draw, fill=gray!15, rounded corners, text width=3.2cm, minimum height=1.6cm] (era1)
{ \textbf{Classical SMC}\\ Robust but discontinuous \\ (High chattering) };
\node[draw, fill=blue!10, rounded corners, text width=3.5cm, minimum height=1.6cm, right=of era1] (era2)
{ \textbf{Higher-Order / Terminal SMC}\\ Finite-time convergence \\ Reduced chattering };
\node[draw, fill=green!10, rounded corners, text width=3.5cm, minimum height=1.6cm, right=of era2] (era3)
{ \textbf{Adaptive and Observer-Based SMC}\\ Online gain tuning \\ Disturbance estimation };
\node[draw, fill=orange!10, rounded corners, text width=3.5cm, minimum height=1.6cm, right=of era3] (era4)
{ \textbf{Intelligent / Data-Driven SMC}\\ NN, RL, and Fuzzy tuning \\ Self-learning control };
\node[draw, fill=yellow!20, rounded corners, text width=3.5cm, minimum height=1.6cm, right=of era4] (era5)
{ \textbf{Next Generation}\\ Hybrid Fractional-Adaptive SMC \\ Energy- and Fault-Aware\\ HIL-validated design };

\draw[->, thick] (era1) -- (era2);
\draw[->, thick] (era2) -- (era3);
\draw[->, thick] (era3) -- (era4);
\draw[->, thick] (era4) -- (era5);

\node[below=0.5cm of era1, font=\footnotesize, text width=3.2cm, align=center]
{1980s–2000s \\ Classical robust control};
\node[below=0.5cm of era2, font=\footnotesize, text width=3.5cm, align=center]
{2000s–2015 \\ Finite-time and higher-order};
\node[below=0.5cm of era3, font=\footnotesize, text width=3.5cm, align=center]
{2015–2020 \\ Adaptive + Observer enhanced};
\node[below=0.5cm of era4, font=\footnotesize, text width=3.5cm, align=center]
{2020–2025 \\ Learning-based hybrid control};
\node[below=0.5cm of era5, font=\footnotesize, text width=3.5cm, align=center]
{Beyond 2025 \\ Intelligent fractional-hybrid frameworks};
\end{tikzpicture}}

\caption{Evolution of Sliding-Mode Control (SMC) strategies for PMSM drives,
showing the shift from discontinuous classical designs to adaptive, observer-based,
and intelligent hybrid controllers. Future development emphasizes fractional-adaptive
structures, energy awareness, and real-time hardware validation.}
\label{fig:smc_roadmap}
\end{figure*}
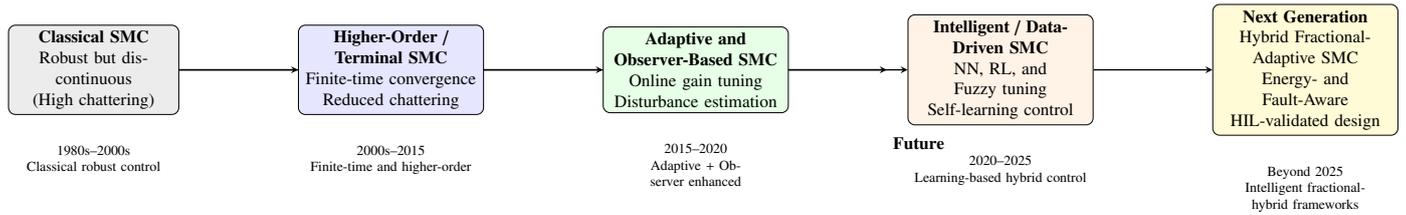

\subsection{Future Research Directions}

Future progress in PMSM-SMC integration will likely revolve around the following strategic themes:

\begin{itemize}
  \item \textbf{Intelligent Adaptation and Self-Learning:}
        By combining SMC and reinforcement learning, fuzzy logic, or neural networks for adaptive gain and surface automation under changing operating conditions.
        These AI-extended SMCs may develop from fixed-parameter controllers to continuously adapting self-organizing controllers.
  \item \textbf{Fractional-Adaptive Hybrid Control:}
        Exploring hybrid controllers that merge fractional-order dynamics with adaptive gain scheduling to exploit the
        memory effect of fractional calculus while avoiding steady-state bias.
  \item \textbf{Energy- and Fault-Aware SMC:}
        Incorporating energy consumption and fault-tolerance metrics into the controller optimization objective, ensuring
        minimal torque ripple and resilience under open-phase or sensor faults.
  \item \textbf{Hardware-in-the-Loop (HIL) Validation:}
        Expanding simulation-only studies into real-time HIL platforms or DSP/FPGA test benches to assess computational
        latency, switching noise, and thermal limits.
  \item \textbf{Benchmark Standardization:}
        Establishing open-source PMSM models and standard disturbance profiles for fair evaluation of emerging SMC designs,
        analogous to how benchmark datasets accelerated progress in AI and robotics.
  \item \textbf{Cross-Domain Integration:}
        Investigating multi-machine coordination, microgrid applications, and vehicular powertrains where SMC principles can
        synergize with model predictive and observer-based control frameworks.
\end{itemize}

\subsection{Concluding Remarks}
The simultaneous development of SMC for PMSM drives shows a broader shift from mathematically elegant but discontinuous controllers to smooth, intelligent, and energy-aware architectures. The technique's ongoing relevance depends on its ability to balance simplicity with robustness and adapt to the efficiency and computing needs of modern electromechanical systems. Future research that includes experimental validation, systematic benchmarking, and intelligent co-design is set to shape the next decade of resilient PMSM control advancements. Figure \ref{fig:smc_roadmap} shows the evolution of the research over the past 25 years and provides a prediction for the future research direction.

\begin{table}[htbp]
\centering
\caption{List of acronyms used throughout the paper for key motor-control and optimization strategies.}
\label{tab:acronyms}
\begin{tabular}{|m{6em}|p{6.5cm}|}
\hline
\textbf{Acronym} & \textbf{Meaning} \\ \hline

\multicolumn{2}{|c|}{\textbf{Motor and Control Fundamentals}} \\ \hline
PMSM & Permanent-Magnet Synchronous Motor \\
FOC  & Field-Oriented Control \\
PID  & Proportional–Integral–Derivative \\ \hline

\multicolumn{2}{|c|}{\textbf{Sliding-Mode Control (SMC) Variants}} \\ \hline
SMC    & Sliding-Mode Control \\
ASMC   & Adaptive Sliding-Mode Control \\
NSMC   & Nonlinear Sliding-Mode Control \\
HOSMC  & Higher-Order Sliding-Mode Control \\
SOSMC  & Second-Order Sliding-Mode Control \\
STSMC  & Super-Twisting Sliding-Mode Control \\
TSMC   & Terminal Sliding-Mode Control \\
NTSMC  & Non-Singular Terminal Sliding-Mode Control \\
FOSMC  & Fractional-Order Sliding-Mode Control \\
FTSMC  & Fast Terminal Sliding-Mode Control \\ \hline

\multicolumn{2}{|c|}{\textbf{Observers and Estimators}} \\ \hline
DOB   & Disturbance Observer \\
NDO   & Nonlinear Disturbance Observer \\
ESO   & Extended State Observer \\
ESDO  & Extended Sliding-Mode DOB \\
FTDO  & Finite-Time Disturbance Observer \\
SMO   & Sliding-Mode Observer \\ \hline

\multicolumn{2}{|c|}{\textbf{Optimization and Intelligent Algorithms}} \\ \hline
GA    & Genetic Algorithm \\
PSO   & Particle Swarm Optimization \\
HHO   & Harris Hawk Optimization \\
RL    & Reinforcement Learning \\
NN    & Neural Network \\ \hline

\multicolumn{2}{|c|}{\textbf{Performance Indices and Auxiliary Terms}} \\ \hline
IAE   & Integral of Absolute Error \\
ISE   & Integral of Squared Error \\
ITAE  & Integral of Time-Weighted Absolute Error \\
ITSE  & Integral of Time-Weighted Squared Error \\
SMRL  & Sliding-Mode Reaching Law \\
HRL   & Hybrid Reaching Law \\
SVPWM & Space-Vector Pulse-Width Modulation \\ \hline

\end{tabular}
\end{table}

\section{Conclusion}\label{sec:conclusion}
This vast literature on Sliding Mode Control for PMSM drives shows a pronounced evolutionary pattern, spurred along by the continuing imperative to overcome the essential trade-off between robustness and the real-world implementation ability limitations of actual actuators. The evolution of the technique, from the Classic SMC to its sophisticated latecomers, constitutes a tale of increasingly elaborate remedies to the essential issue of chattering, as well as attempts to optimize further performance indices such as convergence rate and steady-state error.

Conventional SMC laid out the fundamental principle of robust control via a discontinuous law, but was marred by the undesirable consequences of chattering. Adaptive SMC presented a reactive remedy by adjusting control effort dynamically. Still, Higher-Order SMC, and notably the Super-Twisting Algorithm, presented a proactive and more conclusive remedy by systematically creating a continuous control signal. Terminal SMC, meanwhile, remedied the asymptotic convergence limitation by incorporating nonlinear manifolds that are sure to provide finite-time stabilization, a key consideration in high-precision, time-critical control. Variants such as Integral SMC addressed the structural liability of the reaching phase. At the same time, Fractional-Order SMC added the principle of system memory to enable more subtle dynamic shaping, but at a considerable price in complexity. The current landscape is controlled by hybrid architectures that enhance SMC with observers, intelligent systems, and optimization algorithms. This tendency indicates a mature realization that the best controller oftentimes is a composite system, in which the strong point of a particular component makes up for the weak point of a second component. The selection of an SMC strategy, thus, is not a quest for a universal "best" algorithm but an application-motivated engineering choice. It relies on a skillful balancing of requirements for performance, like tracking accuracy, disturbance rejection, and convergence rate, against implementation constraints, like accessible computing capacity, design time, and cost of the system. As continued research aims to overcome the outstanding challenges of real-time implementation, formal verification, and fault tolerance, SMC is well-positioned to become a crucial and dynamic instrument for facilitating the next family of high-performance electromechanical systems.

\section*{Funding Declaration}
No funding was received for this work.

\bibliographystyle{elsarticle-num} 
\bibliography{references}
\end{document}